\documentclass{aa}  

\usepackage{graphicx}
\usepackage{txfonts}
\usepackage{booktabs}
\usepackage{gensymb}
\usepackage{caption}
\usepackage{siunitx}
\usepackage{hyperref}
\usepackage{natbib}
\usepackage{placeins}
\usepackage{float}
\hypersetup{
    colorlinks,
    citecolor=blue,
    filecolor=magenta,      
    linkcolor=blue,
    }

\titlerunning{Molecular ratios as AGN diagnostics}
\begin{document}

   \title{Understanding if molecular ratios can be used as diagnostics of AGN and starburst activity: The case of NGC 1068.}

   \subtitle{}

   \author{J. Butterworth
          \inst{1},
          J. Holdship\inst{1,2}
          S. Viti\inst{1,2}
          and S. Garc\'ia-Burillo\inst{3}}

   \institute{Leiden Observatory, Leiden University, PO Box 9513, NL-2300 RA Leiden, the Netherlands \and Department of Physics and Astronomy, University College London, Gower St., London, WC1E 6BT, UK \and Observatorio Astron\'omico Nacional (OAN-IGN)-Observatorio de Madrid, Alfonso XII, 3, E-28014 Madrid, Spain\\
              \email{butterworth@strw.leidenuniv.nl, viti@strw.leidenuniv.nl, holdship@strw.leidenuniv.nl,s.gburillo@oan.es}
             }

   \date{Received 22 July 2022; accepted 9 September 2022}

 
  \abstract
   {Molecular line ratios, such as HCN(1-0)/HCO$^+$(1-0) and HCN(4-3)/CS(7-6), are routinely used to identify active galactic nuclei (AGN) activity in galaxies. Such ratios are, however, hard to interpret as they are highly dependent on the physics and energetics of the gas, and hence can seldom be used as a unique, unambiguous diagnostic.  } 
   {We used the composite galaxy NGC 1068 as a `laboratory' to investigate whether  molecular line ratios between HCN, HCO$^+$, and CS are useful tracers of AGN-dominated gas and  determine the origin of the differences in such ratios across different types of gas. Such a determination will enable a more rigorous use of such ratios. }
   {First, we empirically examined the aforementioned ratios at different angular resolutions to quantify correlations. We then used local thermodynamic equilibrium (LTE) and non-LTE analyses coupled with Markov chain Monte Carlo (MCMC) sampling in order to determine the origin of the underlying differences in ratios. }
   {
   We propose that at high spatial resolution (< 50 pc) the HCN(4-3)/CS(2-1) is a reliable tracer of AGN activity. We also find that the variations in ratios are not a consequence of different densities or temperature but of different fractional abundances, yielding to the important result that it is essential to consider the chemical processes at play when drawing conclusions from radiative transfer calculations.}
   {From analyses at varying spatial scales, we find that previously proposed molecular line ratios, as well as a new one, have varying levels of consistency. 
   We also determine from an investigation of radiative transfer modelling of our data that it is essential to consider the chemistry of the species when reaching conclusions from radiative transfer calculations. }

   \keywords{Interstellar medium (ISM): molecules, galaxies: active – Seyfert - starburst - ISM, astrochemistry.
               }

   \maketitle
%

\section{Introduction}
\label{sec:Intro}


The gas and dust present  in the interstellar medium (ISM) within galaxies is not homogeneous; star formation, supernovae events, and active galactic nuclei (AGN) activity   may all greatly alter the ISM \citep[see e.g.][]{Meijerink_2005,Bayet_2009,2014Watanabe}. In particular, recent studies of  nearby external galaxies have shown that the molecular ISM  varies at a parsec, as well as at kiloparsec scales, with evidence of different gas components being traced by different molecular species or rotational transitions  \citep{Scourfield_2020}.
The non-homogeneity of the ISM has been observed in galaxies of different types, such as starburst (SB) galaxies \citep{Meier_2015}, AGN-dominated galaxies \citep{Audibert_2019}, and ultra-luminous infrared galaxies (ULIRGs) \citep{Imanishi_2018}. 

\par Molecular line ratio diagnostics are often used to investigate the physics and chemistry of the ISM in all of these  environments.  For example, as the gas chemistry located in the central, nuclear regions of galaxies is believed to be dominated by X-rays produced by the AGN\textbf{, in so-called X-ray dominated regions (XDR),} the molecular content of the ISM surrounding such nuclei will greatly differ from that in starburst regions \citep{Usero_2004,Garcia_Burillo_2010}. Hence, line ratios of specific molecules have been proposed as indicators of certain energetic or physical processes,  for example HCN/HCO$^+$ as a tracer of AGNs \citep{Leonen2007}, HCN/HNC as a mechanical heating tracer \citep{Hacar2020}, and HCN/CO as a  density tracer \citep{Leroy2017}. In particular,  the `submillimeter-HCN diagram', first proposed in \citep{Izumi2013} and later expanded upon in \citep{Izumi2016}, is a very notable example of the use of molecular line ratios as a probe  of AGN-galaxies; this diagram  makes use of two line ratios, HCN(4-3)/HCO$^+$(4-3) and HCN(4-3)/CS(7-6), where all of the molecules involved are considered tracers  of dense gas. \cite{Izumi2016} observed a clear trend that AGNs, including the Seyfert composite galaxy NGC 1068, tend to show higher HCN/HCO$^+$ and/or HCN/CS than in SB galaxies as long as the observations were at high enough spatial resolutions to separate energetically discrete regions. 
\cite{Izumi2016} propose a scenario where it is the high temperature that is responsible for the HCN enhancement, whereby neutral-neutral reactions with high reaction barriers are enhanced \citep{Harada_2010}, thus leading to the possible enhancement of HCN and the depletion of HCO$^+$ via newly available formation and destruction paths, respectively. However, while of course  higher gas temperatures are expected in AGN-dominated regions, these are not unique to these environments, as starburst regions and/or regions where outflows dominate can also harbour high enough temperatures for such enhancement to occur.  Additionally, the higher temperatures could increase HCN excitation, relative to HCO$^+$ and CS, without necessarily changing their relative abundances \citep{Imanishi_2018}. Finally, infrared radiative pumping is also a possible explanation of the HCN intensity enhancement relative to HCO$^+$ and CS. Infrared pumping is a result of the emission of 14 $\mu$m infrared photons due to the presence of hot dust around AGN. These photons vibrationally excite HCN to the $\nu_2=1$ state. Upon de-exciting from this state back to the vibrational ground state, $\nu=0$, the HCN line intensities are thus pumped to higher fluxes \citep{Imanishi_2018}. However, we note that it is also not unlikely that the  12 $\mu$m infrared photons can similarly vibrationally excite HCO$^+$, thus nullifying the extent of this effect \citep{Imanishi_2016b}.

\par In fact, while  \cite{Miyamoto_2017} observed a similar variation consistent with the `submillimeter-HCN' diagram between the circumnuclear disc (CND) and the star-forming ring of the Seyfert 2 galaxy NGC 613, indicating that the enhancement of HCN may indeed be more prominent in AGN-dominated gas, some studies have observed HCN/HCO$^+$ enhancements in SB galaxies similar to those observed in AGNs \citep{Harada_2018,Konig_2018}. 
Furthermore, a statistical study  by \cite{Privon_2020}  surveying a sample of 58 local luminous infrared galaxies and ULIRGs, concluded that an enhancement in the HCN/HCO$^+$(1-0) line ratio could not be shown to be correlated to the AGN activity. They also concluded that HCN/HCO$^+$ ratios are not a dependable method for finding AGN regions in galaxies.
Finally, from a chemical point of view the HCN/HCO$^+$ ratio  has be shown to be highly dependent on numerous factors, such as density, temperatures, and radiation fields, and cannot alone be used as a unique diagnostic of AGN-dominated galaxies \citep{Viti_2017}. 

It is clear that for a meaningful analysis of molecular line ratios, one first needs to know whether there are there unique molecular ratios at a specific spatial resolution that trace distinct types of gas and energetics associated with  AGN- and SB-dominated galaxies.
Additionally, one needs to know if molecular ratios obtained at different spatial resolutions yield different trends. 

\par In order to obtain these answers, we performed a multi-scale investigation into the use of molecular line ratios as tracers, particularly of AGN versus SB activity, by using the galaxy NGC~1068  as a `laboratory'. NGC 1068 is a Seyfert 2 barred spiral galaxy; it is also the archetypal composite AGN/SB galaxy. Its close proximity ($\sim$14 Mpc) and significant brightness ($L_{Bol}$ $\approx$ 2.5 -- 3.0 $10^{11}L_{\odot}$ ) \citep{Bland-Hawthorn_1997,Bock_2000} enable studies at many different resolutions, down to parsec scales, with 1" corresponding to $\sim$70pc for reference \citep{2014_Garcia_Burillo,2016_Garcia_Burillo,2019_Garcia_Burillo,Viti2014, Gallimore_2016,Imanishi_2018b,Impellizzeri_2019,Imanishi_2020}.
 In Sect. \ref{sec:Data} we provide a summary of the   data collated for this study. Section \ref{sec:ratios} presents an analysis of various molecular ratios at both high and low spatial resolution. In Sect. \ref{sec:Propanalysis} we perform an LTE and non-LTE analysis of the line intensities used for the ratios in Sect. \ref{sec:ratios}, in order to determine whether such an analysis is consistent with the trends implied by the ratios. We summarise our findings in Sect. \ref{sec:Conclusion}.

\section{Data}
\label{sec:Data}

Despite the plethora of available published data focused on NGC 1068, within this study we  focus on the dense gas tracers: HCN, HCO$^+$, and CS. This focus is in an effort to validate previously proposed  ratios and diagrams rather than to suggest new ones. We therefore searched published data for observations of these molecules in NGC 1068 and a summary of all of the data utilised within this study is provided in Table \ref{Tab:Archival_Data}. Observations from multiple instruments were used, including interferometric observations from  Atacama Large Millimetre/submillimetre Array (ALMA) and Plateau de Bure Interferometer (PdBI), and single dish observations from the 15m diameter James Clerk Maxwell Telescope (JCMT).



\begin{table*}[]

\centering
\begin{tabular}{ccccc}
\hline
\textbf{Transition} & \textbf{Spatial resolution} & \textbf{Average spatial resolution (pc)} & \textbf{Instrument} & \textbf{Study}\\ \hline
\multicolumn{5}{c}{HCN} \\ \hline
  HCN(1-0)         &  $0.7'' \times    0.5''$ (49 pc $\times$ 35 pc)             &       42                         &  ALMA Band 3   &  1     \\
  HCN(1-0)         &  $2'' \times    0.8''$ (140 pc $\times$ 56 pc)             &        98                         &  PdBI   &  2      \\
    HCN(4-3)         &  $0.6'' \times    0.5''$ (42 pc $\times$ 35 pc)             &       39                         &  ALMA Band 7   &  3,4      \\
 HCN(4-3)         &  14''             &       980                         &  JCMT 15m   &  5      \\
  HCN(4-3)         &  14''             &       980                         &  JCMT 15m   &  6      \\
  \hline
\multicolumn{5}{c}{HCO$^+$} \\ \hline
  HCO$^+$(1-0)         &  $0.7'' \times    0.5''$ (49 pc $\times$ 35 pc)             &       42                         &  ALMA Band 3   &  1      \\
  HCO$^+$(1-0)         &  $2'' \times    0.8''$ (140 pc $\times$ 56 pc)             &        98                         &  PdBI   &  2      \\
  HCO$^+$(4-3)         &  $0.6'' \times    0.5''$ (42 pc $\times$ 35 pc)             &       39                         &  ALMA Band 7   &  3,4      \\
   HCO$^+$(4-3)         &  14''             &       980                         &  JCMT 15m   &  5      \\
  HCO$^+$(4-3)         &  14''             &       980                         &  JCMT 15m   &  6      \\ \hline
  \multicolumn{5}{c}{CS} \\ \hline
  
    CS(2-1)       &   $0.75'' \times    0.51''$ ($\sim53$ pc $\times$ $\sim36$ pc)                 &         $\sim44$                        &    ALMA Band 3 & 7 \\
        CS(2-1)       &   $4.2'' \times    2.4''$ ($294$ pc $\times$ $168$ pc)                 &         231                        &    ALMA Band 3 & 8 \\
CS(2-1)       &   $6.6'' \times    3.2''$ ($464$ pc $\times$ $224$ pc)                 &         343                        &    ALMA Band 3 & 9 \\
CS(7-6)         &  $0.6'' \times    0.5''$ (42 pc $\times$ 35 pc)             &       39                         &  ALMA Band 7   &  3,4,7      \\
CS(7-6)         &  $4.2'' \times    2.4''$ ($294$ pc $\times$ $168$ pc)                 &         231                         &  ALMA Band 7   &  10      \\
CS(7-6)         &  14''             &       980                         &  JCMT 15m   &  11      \\
        \hline
\end{tabular}

\caption{Studies, 1 - This study/\citep{Sanchez_Garcia_2022}, 2 - \cite{GarciaBurillo_S_2008}, 3 - \cite{2014_Garcia_Burillo}, 4 - \cite{Viti2014}, 5 - \cite{Perez_Beaupuits1009}, 6 - \cite{Tan2018}, 7 - \cite{Scourfield_2020}, 8 - \cite{Takano_2014}, 9 - \cite{Tacconi1997}, 10 - \cite{Nakajima2015}, 11 - \cite{Bayet_2009}.}
\label{Tab:Archival_Data}
\end{table*}

To supplement these data,
we also re-analysed previously published data \citep{Sanchez_Garcia_2022} of  HCN (1-0) and HCO$^+$ (1-0) emission observations. Both lines were observed with ALMA's Band 3 receiver during cycle 2 (project-ID: 2013.1.00055.S; PI: S. Garc\'{ı}a-Burillo).

The phase tracking centre was set to $\alpha_{2000} = 02^\text{h}42^\text{m}40.771^\text{s}$, $\delta_{2000}=-00\degree00'47.94"$ (J2000 reference system, as used throughout the paper) in each case, the position of the galaxy’s centre in the SIMBAD Astronomical Database, from the Two-Micron All-Sky Survey (2MASS) \cite{Skrutskie}. This is offset relative to the galaxy AGN at $\alpha_{2000} = 02^\text{h}42^\text{m}40.710^\text{s}$, $\delta_{2000}=-00\degree00'47.94"$ by $<1"$, and corresponds to a peak in CO emission \citep{2014_Garcia_Burillo}.
Initial reduction of the data was carried out using the ALMA reduction package {\tt CASA} \citep{CasaMcmullin}, and then exported to PYTHON for plotting, making use of the {\tt Matplotlib and Astropy}. The 1 $\sigma$ threshold was determined in {\tt CASA} by calculating the rms of the signal over an area towards the centre of the galaxy in a channel with no line detection.
The angular resolution obtained using uniform weighting
was $\sim0.7" \times 0.5"$ at a position angle of $-86\degree$ in the data cube. The conversion factor between Jy beam$^{-1}$ and K is 32 K Jy$^{-1}$ beam.

\subsection{Moment 0 maps}

\begin{figure*}[h]
\centering
\includegraphics[width = \textwidth]{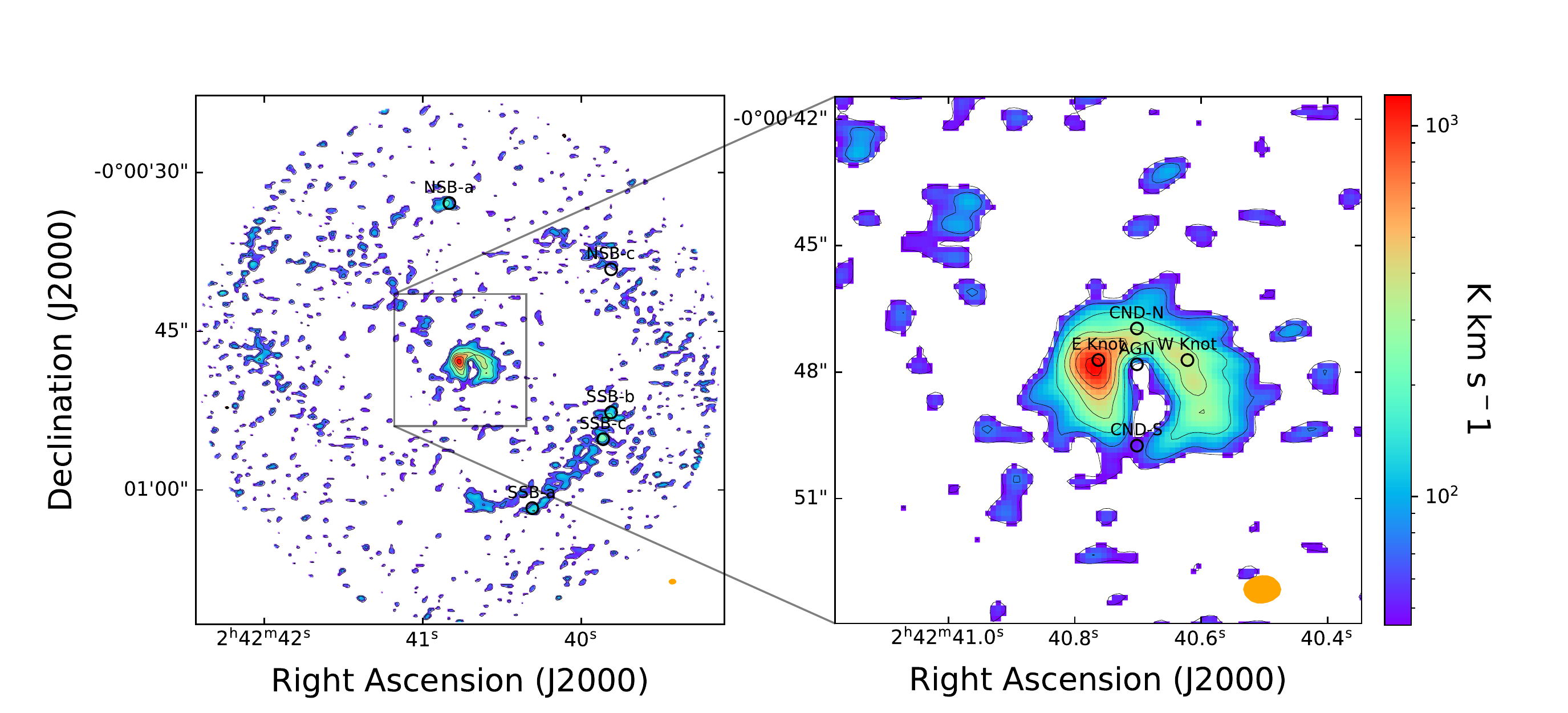}

\caption{HCO$^+$ J(1-0) velocity-integrated moment map. The beam size is shown by the orange ellipse. The lowest contour displayed is 3$\sigma$, where $\sigma=$16 K km s$-1$, with the following contours corresponding to 5$\sigma$, 10$\sigma$, 20$\sigma$, 30$\sigma$, 45$\sigma$, and 70$\sigma$.}
\label{fig:HCO$^+$}
\end{figure*}

\begin{figure*}[h]
\centering
\includegraphics[width = \textwidth]{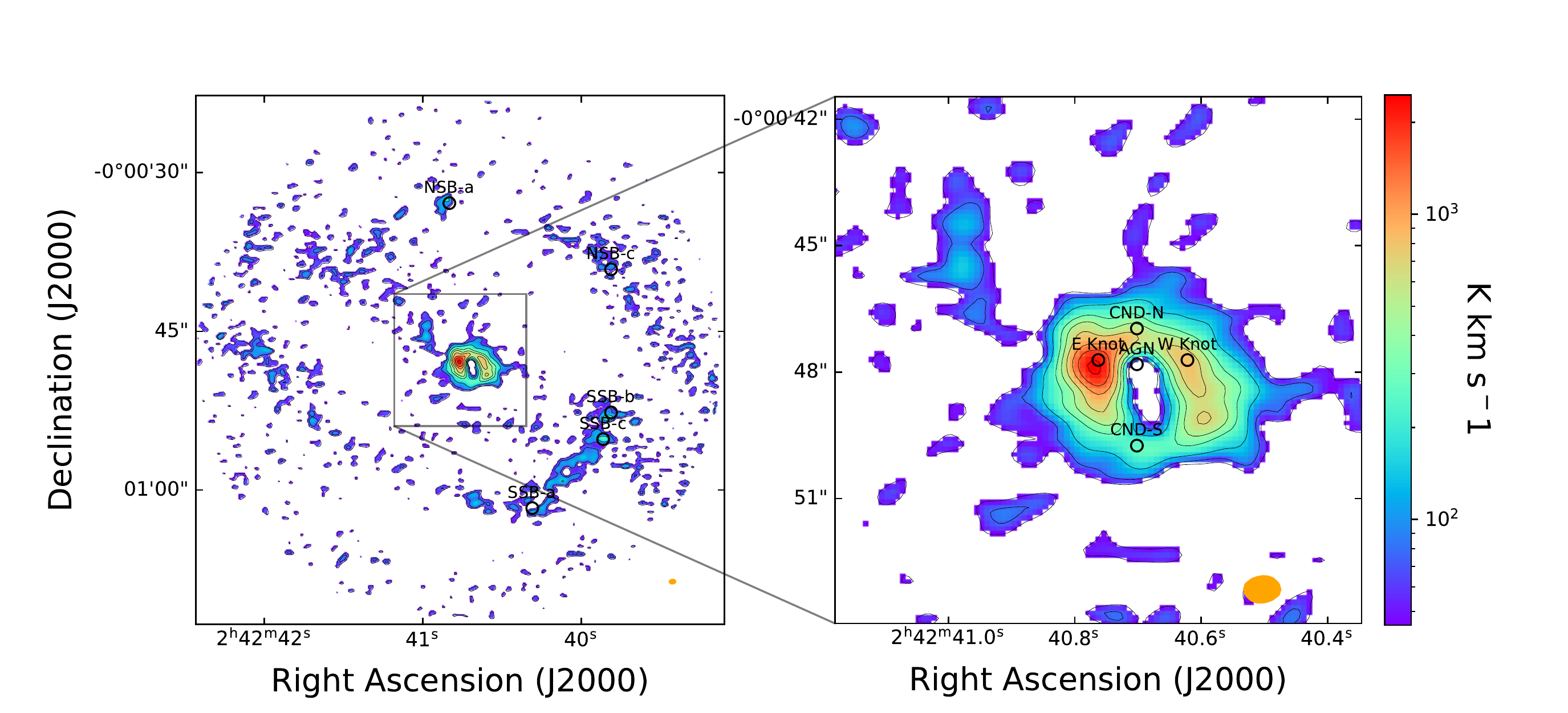}
\caption{HCN J(1-0) velocity-integrated moment map. The beam size is shown by the orange ellipse. The lowest contour displayed is 3$\sigma$, where $\sigma=$16 K km s$-1$, with the following contours corresponding to 5$\sigma$, 10$\sigma$, 20$\sigma$, 30$\sigma$, 45$\sigma$, 70$\sigma$, 100$\sigma$, and 150$\sigma$. }
\label{fig:HCN}
\end{figure*}

Figures \ref{fig:HCO$^+$} and \ref{fig:HCN} show the velocity integrated maps at the original spatial resolution for HCO$^+$(1-0) and HCN(1-0), respectively. As can be seen in these two figures, HCN(1-0) is observed to be significantly more prominent in the CND regions than the equivalent HCO$^+$(1-0) emission. 

The structure of the CND ring as observed in previous observations \citep{2014_Garcia_Burillo,2019_Garcia_Burillo} at similarly  high resolution in CO, CS, and higher J-transitions of HCN and HCO$^+$, is observed in these intensity maps, with the east and west knots showing particularly prominent emission in these lines.
The two knots are connected by `bridges' of emission, within which the CND-N and CND-S regions are located. Emission in these `bridges' is seen to be significantly more prominent in HCN(1-0) rather than HCO$^+$(1-0). Within the SB ring regions, however, HCO$^+$ is observed to be most prominent, the notable exception being NSB-c at which significant HCN emission is seen to dominate over the respective HCO$^+$ emission; the implications of this are further elaborated upon in the following section. As both lines were taken from the same data set at the same resolution, their respective rms are both approximately $\sigma=$16 K km s$^{-1}$ at this resolution.

Moment 0 maps for the remaining high-resolution data can be found in the following papers: the CS(2-1) moment 0 map in Fig. A1 of \cite{Scourfield_2020}; CS(7-6) in Fig. A3 of \cite{Scourfield_2020}; HCO$^+$(4-3)  in Fig. 8 of \cite{2014_Garcia_Burillo}; a zoomed-in moment 0 map of HCN(4-3) is  presented  in Fig. 8 of \cite{2014_Garcia_Burillo}, with a zoomed-out equivalent also available in the appendix of this paper in Section \ref{appsubsec:HCN_43}, with an accompanying HCO+(4-3) plot in Section \ref{appsubsec:HCOP_43}. 
\begin{table}[]
\caption{Names and coordinates of the sub-regions of interest across the CND and the SB ring.} 
\begin{tabular}{lcc}\hline
\textbf{Sub-region name} & \textbf{RA}                                 & \multicolumn{1}{c}{\textbf{Declination}}    \\ \hline
E Knot          & $02^{h}42^{m}40^{s}.771$ & $-00^{\circ}00'47''.84$ \\ \hline
W Knot          & $02^{h}42^{m}40^{s}.630$ & $-00^{\circ}00'47''.84$ \\ \hline
AGN             & $02^{h}42^{m}40^{s}.710$ & $-00^{\circ}00'47''.94$ \\ \hline
CND-N           & $02^{h}42^{m}40^{s}.710$  & $-00^{\circ}00'47''.09$ \\ \hline
CND-S           & $02^{h}42^{m}40^{s}.710$  & $-00^{\circ}00'49''.87$ \\ \hline
NSB-a           & $02^{h}42^{m}40^{s}.840$ & $-00^{\circ}00'33''.00$ \\ \hline
NSB-c           & $02^{h}42^{m}39^{s}.820$ & $-00^{\circ}00'39''.26$ \\ \hline
SSB-a           & $02^{h}42^{m}40^{s}.317$ & $-00^{\circ}01'01''.84$ \\ \hline
SSB-b           & $02^{h}42^{m}39^{s}.820$ & $-00^{\circ}00'52''.79$ \\ \hline
SSB-c           & $02^{h}42^{m}39^{s}.870$ & $-00^{\circ}00'55''.32$ \\ \hline
\end{tabular}
\label{tab:Regions}
\end{table}

\begin{figure*}[h]
\centering
\includegraphics[width = \textwidth]{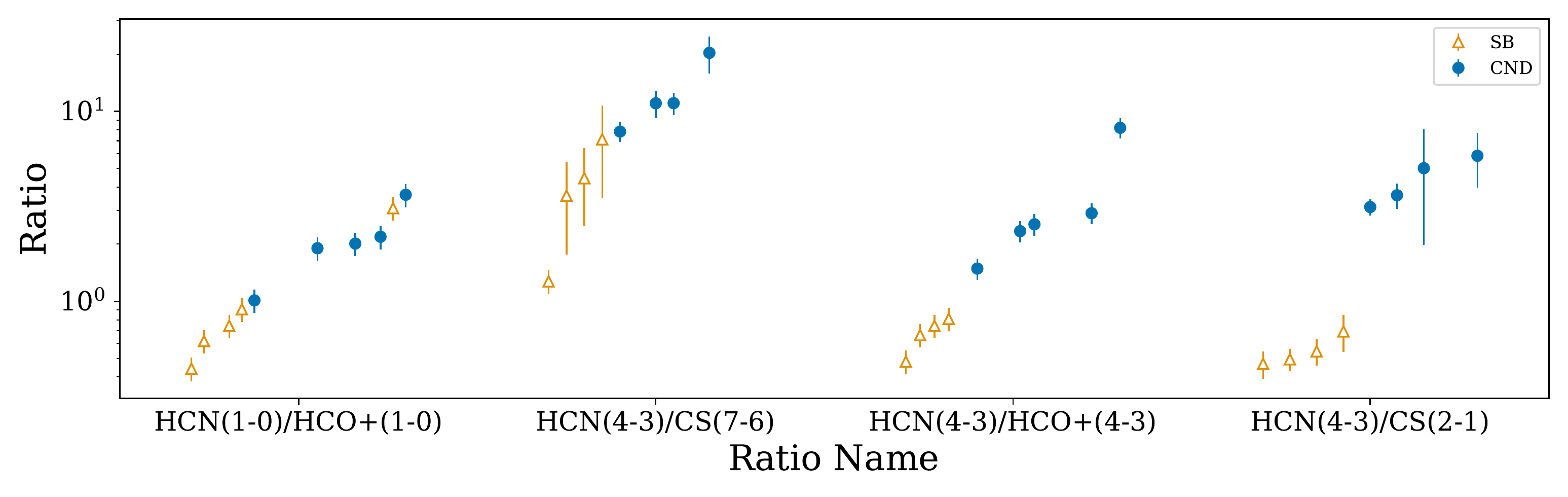}
\caption{The high-resolution ($\sim40$ pc scale) observations of the four ratios investigated in this paper, split into regions located in the CND and the SB ring. Values and uncertainties of the ratio values plotted here are given in Table \ref{tab:Ratios}.}
\label{fig:High_ResSwarm}
\end{figure*} 

\begin{figure}[h]
\centering
\includegraphics[width = 0.5\textwidth]{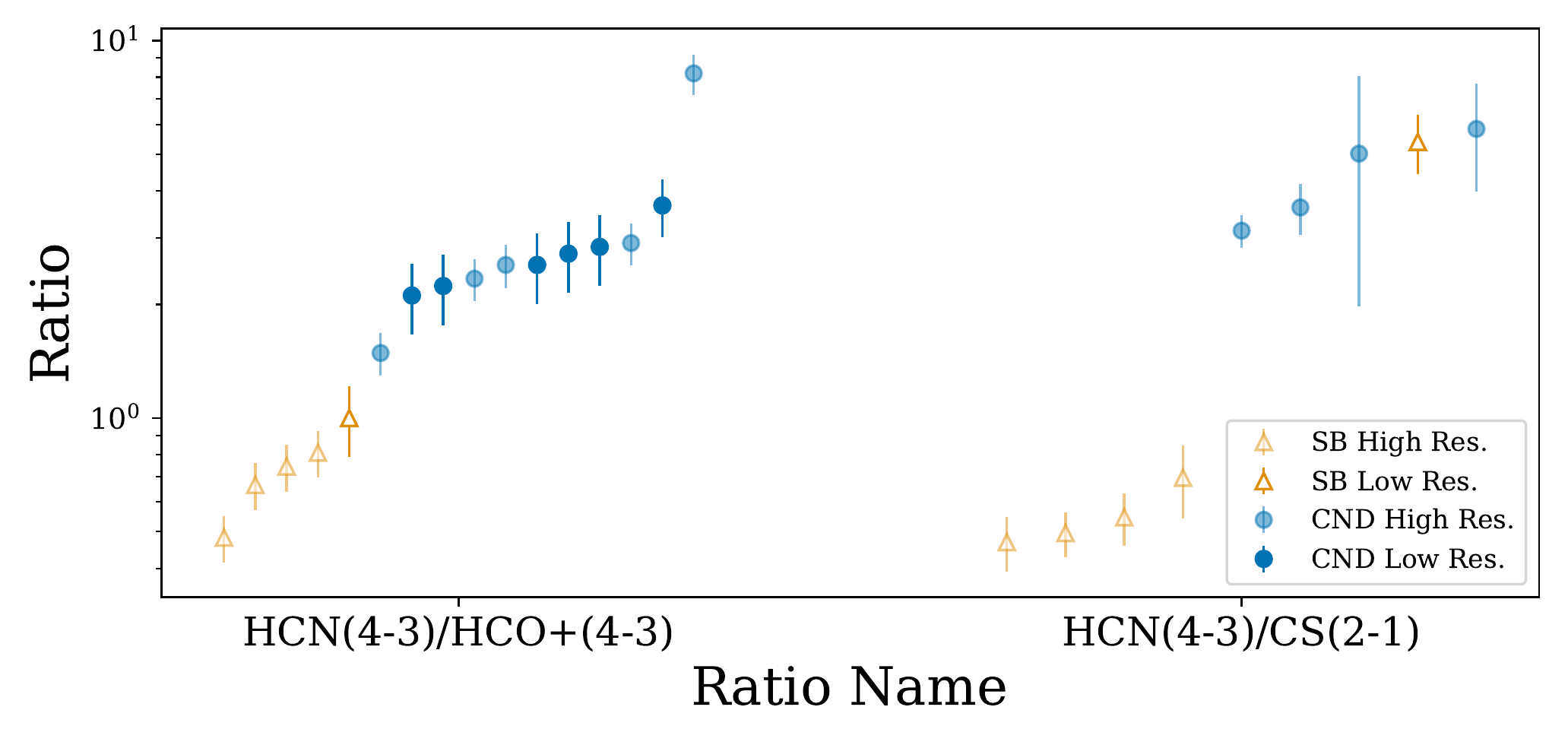}
\caption{The data from Fig. \ref{fig:High_ResSwarm} with the addition of lower-resolution data (see Table \ref{Tab:Archival_Data}) for the two ratios that showed reasonable distinction between the CND and the SB ring regions.}
\label{fig:Swarm_All}
\end{figure} 

\begin{table}[h!]
\caption{The values and uncertainties of the ratio values for the respective ratios displayed in Figs. \ref{fig:High_ResSwarm}, \ref{fig:Swarm_All}, and \ref{fig:HCN43_CS76}, with the corresponding sub-regions given for each ratio value. Data used in Fig. 5 are added for clarity.}
\centering
\begin{tabular}{|ccc|}

\hline
\multicolumn{1}{|c|}{\textbf{Ratio}} & \multicolumn{1}{c|}{\textbf{Ratio error}}  & \textbf{Location} \\ \hline
\multicolumn{3}{|c|}{\textbf{High-resolution ($\sim$40 pc) HCN(1-0)/HCO$^+$(1-0)}}                 \\ \hline
\multicolumn{1}{|c|}{1.01}   & \multicolumn{1}{c|}{0.14}         & AGN      \\ \hline
\multicolumn{1}{|c|}{1.9}    & \multicolumn{1}{c|}{0.27}       & CND-N    \\ \hline
\multicolumn{1}{|c|}{3.64}   & \multicolumn{1}{c|}{0.51}         & CND-S    \\ \hline
\multicolumn{1}{|c|}{2.01}   & \multicolumn{1}{c|}{0.28}      & E Knot   \\ \hline
\multicolumn{1}{|c|}{0.62}   & \multicolumn{1}{c|}{0.09}       & NSB-a    \\ \hline
\multicolumn{1}{|c|}{3.09}   & \multicolumn{1}{c|}{0.44}        & NSB-c    \\ \hline
\multicolumn{1}{|c|}{0.44}   & \multicolumn{1}{c|}{0.06}      & SSB-a       \\ \hline
\multicolumn{1}{|c|}{0.74}   & \multicolumn{1}{c|}{0.11}       & SSB-b    \\ \hline
\multicolumn{1}{|c|}{0.91}   & \multicolumn{1}{c|}{0.13}     & SSB-c    \\ \hline
\multicolumn{1}{|c|}{2.19}   & \multicolumn{1}{c|}{0.31}        & W Knot   \\ \hline
\multicolumn{3}{|c|}{\textbf{High-resolution ($\sim$40 pc) HCN(4-3)/CS(2-1)}}               \\ \hline
\multicolumn{1}{|c|}{5.83}   & \multicolumn{1}{c|}{1.86}      & CND-N    \\ \hline
\multicolumn{1}{|c|}{5.02}   & \multicolumn{1}{c|}{3.04}      & CND-S    \\ \hline
\multicolumn{1}{|c|}{3.14}   & \multicolumn{1}{c|}{0.31}        & E Knot   \\ \hline
\multicolumn{1}{|c|}{0.5}    & \multicolumn{1}{c|}{0.07}       & NSB-a    \\ \hline
\multicolumn{1}{|c|}{0.55}   & \multicolumn{1}{c|}{0.09}      & SSB-a       \\ \hline
\multicolumn{1}{|c|}{0.47}   & \multicolumn{1}{c|}{0.08}       & SSB-b    \\ \hline
\multicolumn{1}{|c|}{0.69}   & \multicolumn{1}{c|}{0.15}        & SSB-c    \\ \hline
\multicolumn{1}{|c|}{3.61}   & \multicolumn{1}{c|}{0.56}      & W Knot   \\ \hline
\multicolumn{3}{|c|}{Low-resolution ($\gtrsim$100 pc) HCN(4-3)/CS(2-1)}                    \\ \hline
\multicolumn{1}{|c|}{5.39}   & \multicolumn{1}{c|}{0.97}        & SSB-a       \\ \hline
\end{tabular}
\label{tab:Ratios}
\end{table}

\begin{table}[h!]
\ContinuedFloat
\caption{Continued.}
\centering
\begin{tabular}{|ccc|}

\hline
\multicolumn{1}{|c|}{\textbf{Ratio}} & \multicolumn{1}{c|}{\textbf{Ratio error}}  & \textbf{Location} \\ \hline
\multicolumn{3}{|c|}{\textbf{High-resolution ($\sim$40 pc) HCN(4-3)/CS(7-6)}}                                                              \\ \hline
\multicolumn{1}{|c|}{11.02}  & \multicolumn{1}{c|}{1.81}       & AGN      \\ \hline
\multicolumn{1}{|c|}{20.3}   & \multicolumn{1}{c|}{4.51}       & CND-N    \\ \hline
\multicolumn{1}{|c|}{7.82}   & \multicolumn{1}{c|}{0.93}         & E Knot   \\ \hline
\multicolumn{1}{|c|}{1.27}   & \multicolumn{1}{c|}{0.18}      & NSB-a    \\ \hline
\multicolumn{1}{|c|}{4.44}   & \multicolumn{1}{c|}{1.96}     & SSB-a       \\ \hline
\multicolumn{1}{|c|}{3.6}    & \multicolumn{1}{c|}{1.83}      & SSB-b    \\ \hline
\multicolumn{1}{|c|}{7.12}   & \multicolumn{1}{c|}{3.63}    & SSB-c    \\ \hline
\multicolumn{1}{|c|}{11.04}  & \multicolumn{1}{c|}{1.46}        & W Knot   \\ \hline
\multicolumn{3}{|c|}{\textbf{Low-resolution ($\gtrsim$100 pc) HCN(4-3)/CS(7-6)}}                                                            \\ \hline
\multicolumn{1}{|c|}{20.92}   & \multicolumn{1}{c|}{4.44}        & AGN      \\ \hline
\multicolumn{1}{|c|}{17.09}   & \multicolumn{1}{c|}{3.62}        & CND-N    \\ \hline
\multicolumn{1}{|c|}{10.97}   & \multicolumn{1}{c|}{2.33}         & CND-S    \\ \hline
\multicolumn{1}{|c|}{77.41}   & \multicolumn{1}{c|}{16.42}     & E Knot   \\ \hline
\multicolumn{1}{|c|}{44.64}   & \multicolumn{1}{c|}{9.47}      & W Knot    \\ \hline
\multicolumn{1}{|c|}{9.93}   & \multicolumn{1}{c|}{3.73}        & AGN ($\sim$1000 pc scale)      \\ \hline
\end{tabular}
\end{table}

\begin{table}[h!]
\ContinuedFloat
\caption{Continued.}
\centering
\begin{tabular}{|ccc|}

\hline
\multicolumn{1}{|c|}{\textbf{Ratio}} & \multicolumn{1}{c|}{\textbf{Ratio error}}  & \textbf{Location} \\ \hline
\multicolumn{3}{|c|}{\textbf{High-resolution ($\sim$40 pc) HCN(4-3)/HCO+(4-3)}}                                                            \\ \hline
\multicolumn{1}{|c|}{1.49}   & \multicolumn{1}{c|}{0.19}        & AGN      \\ \hline
\multicolumn{1}{|c|}{2.34}   & \multicolumn{1}{c|}{0.3}        & CND-N    \\ \hline
\multicolumn{1}{|c|}{8.19}   & \multicolumn{1}{c|}{1.01}         & CND-S    \\ \hline
\multicolumn{1}{|c|}{2.54}   & \multicolumn{1}{c|}{0.33}     & E Knot   \\ \hline
\multicolumn{1}{|c|}{0.48}   & \multicolumn{1}{c|}{0.07}      & NSB-a    \\ \hline
\multicolumn{1}{|c|}{0.74}   & \multicolumn{1}{c|}{0.11}       & SSB-a       \\ \hline
\multicolumn{1}{|c|}{0.81}   & \multicolumn{1}{c|}{0.11}       & SSB-b    \\ \hline
\multicolumn{1}{|c|}{0.67}   & \multicolumn{1}{c|}{0.09}      & SSB-c    \\ \hline
\multicolumn{1}{|c|}{2.91}   & \multicolumn{1}{c|}{0.37}       & W Knot   \\ \hline
\multicolumn{3}{|c|}{Low-resolution ($\gtrsim$100 pc) HCN(4-3)/HCO+(4-3)}                                      \\ \hline
\multicolumn{1}{|c|}{2.11}   & \multicolumn{1}{c|}{0.45}        & AGN      \\ \hline

\multicolumn{1}{|c|}{2.24}   & \multicolumn{1}{c|}{0.47}       & CND-N    \\ \hline
\multicolumn{1}{|c|}{2.73}   & \multicolumn{1}{c|}{0.58}     & CND-S    \\ \hline
\multicolumn{1}{|c|}{2.55}   & \multicolumn{1}{c|}{0.54}     & E Knot   \\ \hline
\multicolumn{1}{|c|}{1.0}    & \multicolumn{1}{c|}{0.21}      & SSB-a       \\ \hline
\multicolumn{1}{|c|}{2.84}   & \multicolumn{1}{c|}{0.6}                & W Knot   \\ \hline
\multicolumn{1}{|c|}{3.66}   & \multicolumn{1}{c|}{0.64}                & AGN ($\sim$1000 pc scale)   \\ \hline
\end{tabular}
\end{table}

\subsection{Comparison between observations}
\label{subsec:obscompare}
The range of transitions from multiple studies will allow us to approach the analysis of NGC 1068 from both a large-scale view, using the low spatial resolution detections of single dish instruments, and a more zoomed-in one, provided to us via interferometric instruments. The different spatial resolutions make it necessary to ensure that only similarly obtained detections are compared. \textbf{This is especially true for the higher-resolution observations that cover only a small area of the galaxy.} For those observations, the variance over even small distances can be large. An example of such small-scale variation in emission can be seen in the CND in Fig~\ref{fig:HCO$^+$}.

We therefore grouped our data in two ways. Firstly, the archival data were grouped by spatial resolution. Within these groupings, there were variations of resolution of up to a factor of two. Therefore, we deconvolved the intensities by assuming the observation with the lowest resolution was the source size and using the equation
\begin{equation}
I_{Source}=\left(\frac{(\theta_{s}^2+\theta_{b}^2)}{\theta_{b}^2}\right)I_{Beam},
\end{equation}
where $I_{Source}$ is the source averaged peak intensity, $\theta_{s}$ is the source size (as defined by the lowest-resolution observation in the category), $\theta_{b}$ is the beam size, and $I_{Beam}$ is the original intensity as defined in the archive or paper. This is similar to the approach taken  in \cite{Kamenetzky2011,Aladro2013,Aladro2015} in order to account for the effect of beam dilution.

The high-resolution data were then further separated by the sub-region on which they were focused. For the low-resolution data, this was not a concern as each observation contained many sub-regions. We list each sub-region in Table \ref{tab:Regions}. The regions in the CND and the SB ring observed within this report are consistent with those observed in \cite{Scourfield_2020}, with the exception of the region known as NSB-b, as none of the ratios investigated within this report have been observed at a consistent resolution for this region. We note that not all sub-regions listed in Table \ref{tab:Regions}, and shown in Figures \ref{fig:HCO$^+$} and \ref{fig:HCN}, have been observed at all resolutions for all molecules. As a result, some ratios in some regions are missing  at certain resolutions (see Fig. \ref{fig:High_ResSwarm}). The region NSB-c possesses fewer observed ratios than the other regions, with just a single high-resolution observation (see Table \ref{tab:Ratios}). As a result it is not included in the \texttt{RADEX} fitting discussed later in this paper.

\section{Molecular line ratios}
\label{sec:ratios}

The combination of archival and new data allows us to investigate molecular line ratios that have been proposed as tracers of different regions as well as to propose a new tracer.
In this section, we take an in-depth look at these ratios across the resolutions and sub-region groupings in order to determine which relationships stand up to scrutiny.

\subsection{HCN/HCO$^+$}

HCN and HCO$^+$ are both dense gas tracers, with HCN(1-0) and HCO$^+$(1-0) having critical densities of $\sim$\SI{e5}{\per\centi\metre\cubed} and $\sim$\SI{5e4}{\per\centi\metre\cubed} (at $\sim$ 10 K, \cite{Shirley}), respectively. The HCN(1-0)/HCO$^+$(1-0) and HCN(4-3)/HCO$^+$(4-3) ratios have both been proposed as tracers of AGN activity \citep{Kohno_2005,Izumi2013}. Therefore, in this section, we discuss each ratio in turn, with a particular focus on how well the ratio separates AGN regions from SB regions. All the ratios for NGC 1068 are presented in Figs. \ref{fig:Swarm_All} and \ref{fig:High_ResSwarm}.

\subsubsection{HCN(1-0)/HCO$^+$(1-0)}
An enhancement of HCN(1-0) relative to HCO$^+$(1-0) or CO(1-0)  has been proposed as an indicator of AGN activity \citep{Imanishi_2007,Krips_2008,Davies_2012}. In fact, \citet{Kohno2001} proposed the HCN diagram; a diagnostic diagram to identify AGN galaxies using this line ratio and the ratio of HCN(1-0)/CO(1-0). 
However, as detailed in the Introduction, some  studies find high HCN(1-0)/HCO$^+$(1-0) ratios in SB galaxies \citep{Costagliola_2011} and low values of this ratio in AGNs \citep{Sani_2012}, which casts doubt on the use of this ratio to trace AGN activity.
In the high-resolution archival data, as well as the new ALMA data we have collated, we typically observe the same distinction in this ratio as in other studies. We find that a HCN(1-0)/HCO$^+$(1-0) ratio of greater than one is located in regions near the AGN, and a ratio less than one is found in those outside of the CND (see Fig. \ref{fig:High_ResSwarm}). Alhough  it must be noted that at the X-ray matched position of the AGN, the sub-region denoted as AGN, the observed ratios are significantly lower than in the other CND regions, which may be consistent with the conclusion reached by \cite{Privon_2020} that HCN(1-0)/HCO$^+$(1-0) does not trace AGN activity directly. We also note that in \cite{2014_Garcia_Burillo} and \cite{Viti2014}, large HCN(1-0)/HCO$^+$(1-0) (as well as the ratio of higher J) was in fact associated with large scale shocks that were associated with the molecular outlfow, rather than XDR chemistry, a scenario also explored in \cite{Garcia_Burillo_2010}. Recent galactic studies, such as the study into the prototypical shock region L1157-B1 in \cite{Lefloch2021}, have shown a marked increase in HCN abundance relative to H$_2$ in these environments, relative to pre-shock gas. 

However, there are significant exceptions. Sub-region NSB-c is not associated with the AGN but has the second largest ratio value. We note, however, that all the lines detected in NSB-c are significantly weaker than those found in the other SB regions.

\subsubsection{HCN(4-3)/HCO$^+$(4-3)}
Similar to the enhancement of HCN relative to HCO$^+$ observed in their respective J=1-0 transitions, \cite{Izumi2013} found a similar trend for the J=4-3 transitions. 
 This ratio has advantages over the J=1-0 ratio as higher-resolution imaging is more easily achievable, it can be observed to higher distances  (up to $z \sim 3$ with ALMA), and the transitions have higher critical densities, resulting in lower contamination by foreground or other extended emission \citep{Izumi2016}. HCN(4-3) and HCO$^+$(4-3) possess high and similar critical densities of $n_{crit,HCN(4-3)}\approx8.5 \times 10^6 \text{cm}^{-3}$ and $n_{crit,\text{HCO}^+(4-3)}\sim10^6$cm$^3$ (at $\sim$ 10 K, \cite{Shirley}), which means that they are more likely to trace the same component of the dense gas in any observed region.

Figure~\ref{fig:High_ResSwarm} seems to imply that this ratio does separate AGN and non-AGN regions in our high-resolution data. Moreover, once low-resolution data are included, the ratio continues to distinguish between regions (Fig. \ref{fig:Swarm_All}), albeit with a smaller gap between the two types of region. 
It must be noted, however, that for region NSB-c, both HCN(4-3) and HCO$^+$(4-3) emission is below the 3 $\sigma$ threshold necessary to derive a reliable intensity ratio. Hence, we cannot be sure whether in NSB-c this ratio follows the same trend as HCN(1-0)/HCO$^+$(1-0).

\subsection{HCN/CS}

CS is also a commonly used dense gas tracer;
CS(1-0) has a critical density of $n_{crit}\sim 10^4$cm$^3$ \citep{Shirley}. \cite{Izumi2013} present the use of CS as opposed to HCO$^+$ as the molecule with which the HCN enhancement should be measured because earlier studies had shown that the CS abundance varies little  between active regions (e.g. near the AGN in the CND) and those in SB regions \citep{Martin_2008,Martin_2009}. For this reason, the HCN/CS ratio has been proposed as a possible starburst and AGN tracer. Under XDR conditions, the CS abundance should be unchanged but the HCN abundance is enhanced, leading to an enhancement in the ratio \citep{Kreiger_2020}.

\subsubsection{HCN(4-3)/CS(7-6)}

\cite{Izumi2013} state that the intensity ratio between HCN(J=4-3) and  CS (J=7-6) was found to be high in AGN galaxies ($> 12.7$). 
HCN(4-3) has an excitation temperature of 42.53K and a critical density of $\approx 10^{7} \text{cm}^{-3}$. CS(7-6) has an excitation temperature of 65.8K and a critical density of $\approx 10^{6} \text{cm}^{-3}$ (at $\sim$10 K, \cite{Shirley}).
 \cite{Audibert_2020} observed results consistent with those proposed by the Izumi papers in NGC 1808. 
Despite this, we find that, even at high spatial resolution, this ratio does not clearly distinguish the CND from the SB regions. Considering the uncertainty in the ratios that we calculated, there is significant overlap between the ratio values at positions in each region. Thus, one would not be able to reliably separate an unusually large SB ratio caused by a random error from a large ratio, due to the CND enhancement.

\begin{figure}
\centering
\includegraphics[width = 0.4\textwidth]{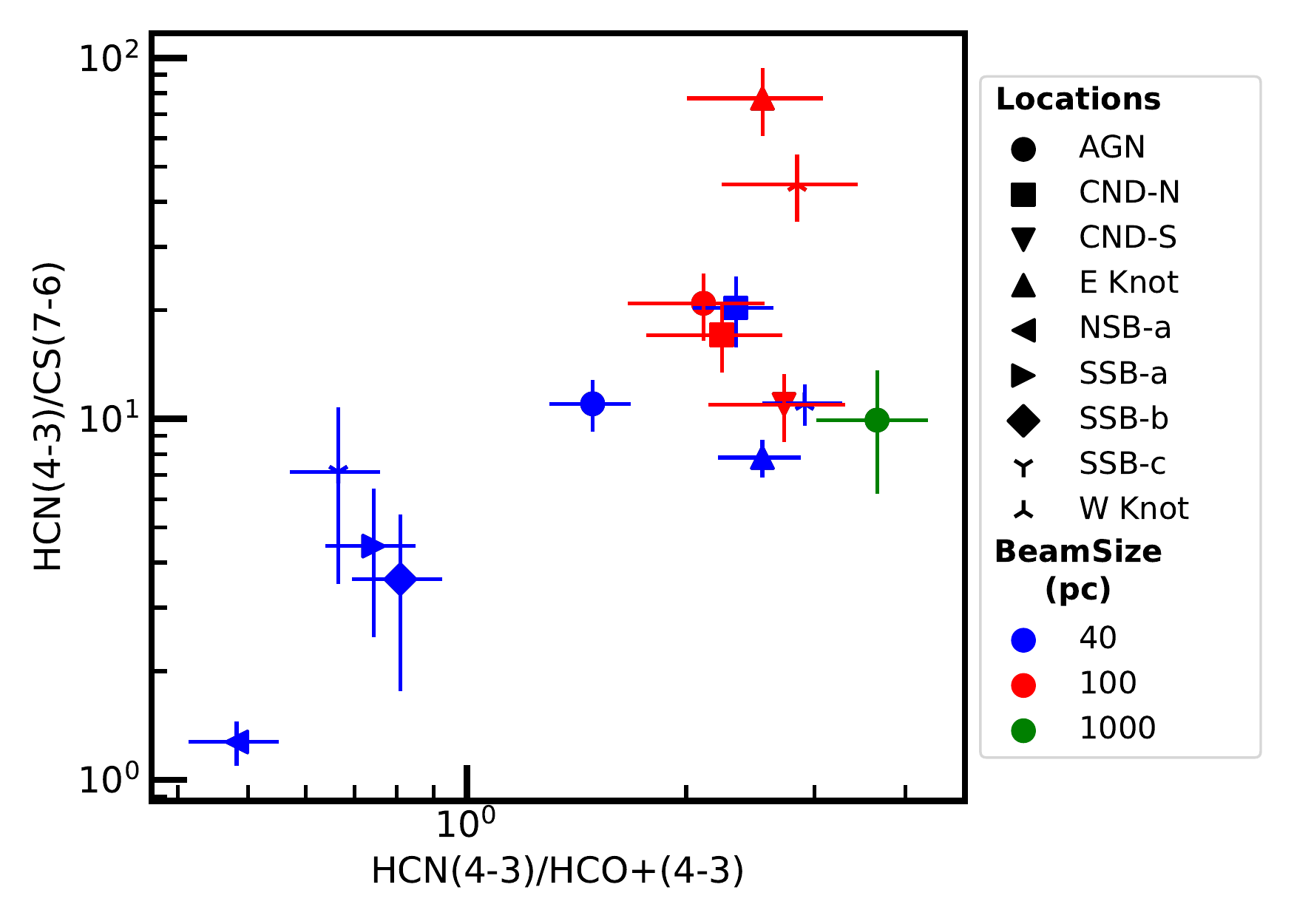}

\caption{Recreation of the sub-millimetre-HCN diagram from \cite{Izumi2016} using various resolution data across various sub-regions in NGC 1068. The shape of each marker denotes the region it corresponds to and the colour denotes the resolution.}
\label{fig:HCN43_CS76}
\end{figure}

One possible reason for the failure of this ratio, given its previous success, is that \cite{Izumi2013} and \cite{Izumi2016} used lower-resolution observations than those used here. Hence, they may trace more extended regions. We indeed also typically find that  the highest values for this ratio are obtained  from the lower-resolution observations in our data. This point is shown quite clearly in Fig. \ref{fig:HCN43_CS76}, which is a recreation of the sub-millimetre-HCN diagram from \cite{Izumi2016} of the HCN(4-3)/HCO$^+$(4-3) and the HCN(4-3)/CS(7-6) ratios.

\subsubsection{HCN(4-3)/CS(2-1)}
We further investigated the HCN(4-3)/CS(2-1) ratio. For NSB-c, as before, we did not have a strong enough detection for HCN(4-3), and as such we could not derive this ratio for this region. Taking a 3$\sigma$ upper limit for the HCN(4-3) in this sub-region, we obtained an upper limit for the HCN(4-3)/CS(2-1) ratio of $\sim$ 1.4, which is significantly lower than the those observed in the CND regions. Considering all the remaining regions, we find that at high spatial resolution, a clear distinction can be made between the CND and SB regions, as shown in Fig. \ref{fig:High_ResSwarm}.  

This is not surprising if we consider the difference in the lines. The HCN (4-3) transition has an upper state energy of 43 K and a critical density of $\sim$ \SI{8.5e6}{\per\centi\metre\cubed}, whilst the CS (2-1) transition has an upper state energy of 7 K and a critical density of  \SI{8e4}{\per\centi\metre\cubed}. Since the two transitions are likely to be excited under very different gas conditions, this ratio should be able to discriminate between SB and CND regions.

However, Fig.~\ref{fig:Swarm_All} shows that once lower-resolution observations are included, this ratio becomes less reliable. In particular, including low-resolution data adds an SB observation that is the second highest observation of this ratio.






\label{sec:Global}

\section{Physical and chemical properties of NGC 1068}
We have found that the ratio between HCN (4-3) and HCO$^+$(4-3) clearly separates SB and AGN regions in our data. We have also found that the ratio between HCN (4-3) and CS (2-1) is reliable at high resolution. However, these are observed correlations with as yet undetermined underlying physical or chemical cause. In this section, we attempt to find that cause so that these ratios can be more robustly used to probe AGN regions.

To achieve this, we used radiative transfer modelling to infer the gas conditions and molecular column densities at each location for each resolution. We were then able to determine whether there are physical differences in excitation conditions between locations, differences in the underlying chemical abundances, or even a combination of the two as a result of shocks\textbf{, or processes involving X-rays or cosmic rays.} If it is the former, these ratios may only distinguish between regions insofar as the gas
temperature and density are typically different. If it is the latter, more detailed chemical modelling
is required to determine the cause of the chemical distinction.

\label{sec:Propanalysis}


\subsection{LTE analysis}
\label{subsec:LTE Analysis}

Using the observed line intensities from the archival data, we were able to estimate the column densities by assuming LTE and by using the same groupings of observations as discussed earlier, devolved to the lowest resolution of that group. The upper state column density of a molecular transition can be approximated to 

\begin{equation}
    N_u = \frac{8\pi k \nu^2W}{hc^3A_{ul}}\left(\frac{\Delta\Omega_a}{\Delta\Omega_s}\right) \frac{\tau}{1 - e^{- \tau}}, 
    \label{eq:uppercol}
\end{equation}
where $W$ is the integrated line intensity, $A_{ul}$ is the  Einstein A-coefficient, and $\tau$ is the optical depth, which we included as a correction factor for saturated lines. $\Omega_a$ and $\Omega_s$ are the solid angles of the antenna and the source, respectively, and together give the beam filling factor. Since we are investigating the regions themselves, we made the assumption that the source completely fills the beam, such that $\frac{\Delta\Omega_a}{\Delta\Omega_s}=1$. 

We could then use these transition column densities to approximate the total column density of the emitting molecule, $N$, by using
\begin{equation}
    N=\frac{N_uZ(T)}{g_ue^{\frac{E_u}{kT}}}
    \label{eq:coldens},
\end{equation}
where $Z(T)$ is the partition function, $g_u$ is the statistical weight of level $u$, and $E_u$ is the excitation energy of level $u$. To compute the partition function, we made use of molecular data from the Cologne Database for Molecular Spectroscopy (CDMS) \citep{CDMS}. 
We combined Eqs.~\ref{eq:uppercol} and~\ref{eq:coldens} to produce an estimate of the column density for each molecule. We used temperatures in the range 50 K to 400 K and computed the column density from each transition independently, with this range of temperatures being based upon previous modelling studies of these regions in \cite{Viti2014} and \cite{Izumi2016}. We set $\tau = 0$ for our lines, following the example of \cite{Krips_2008}. By following this procedure, we obtained a broad range of possible values for the column densities, which act as limits for the grid of radiative transfer models used to obtain non-LTE results in the next section. Table \ref{Tab:ColumnDens} shows the minimum and maximum LTE column density results for each molecule.



\begin{table}[]
\caption{Lower and upper limits of column density calculated for each molecule assuming LTE.}
\centering
\begin{tabular}{ccc}
\toprule
\textbf{Molecule} & \textbf{$N_{min}$ (cm$^{-2}$)} & \textbf{$N_{max}$ (cm$^{-2}$)} \\
\midrule
CS & 3.34E+12 & 2.44E+16 \\
HCO+ & 1.73E+11 & 3.43E+16 \\ 
HCN & 2.62E+13 & 7.38E+17 \\
\bottomrule
\end{tabular}

\label{Tab:ColumnDens}
\end{table}




\begin{table}[]
\caption{\textbf{The range of physical parameters over which the \texttt{RADEX} models were computed. $N_{min}$ and $N_{max}$ are the respective minimum and maximum column densities for each molecule, as shown in Table \ref{Tab:ColumnDens}.}}
\centering
\begin{tabular}{cccc}
\toprule
Parameter  & Lower  & Upper  & Discrete  \\
                            (unit)      &  limit                            &  limit                            &      Values                             \\ \midrule
$n_{H_2}$ (cm$^{-3}$)               & $10^3$                       & $10^7$                       & 50                                \\ \hline
$T_K$ (K)                         & 50                           & 400                          & 50                                \\ \hline
$N_{Mol}$ (cm$^{-2}$)               & $N_{min}$                    & $N_{max}$                    & 50                                \\ \hline
$\Delta v$ (km s$^-1$)                       & 50                           & 150                          & 3                                 \\ \bottomrule
\end{tabular}
\label{tab:Grid_Range}
\end{table}
\begin{table*}[]

\centering
\caption{Median, 16th, and 84th percentile parameter value results from the MCMC chains resulting from fitting to the observed ratios.  }

\begin{tabular}{|c|c|c|c|c|c|c|c|}
\hline
\textbf{$\sim$Beam size (pc)}  & \textbf{Region} & \textbf{Log($n_H$)}            & \textbf{Temperature (K)}            & \textbf{Linewidth}                & \textbf{Log($N_{HCN}$)}           & \textbf{Log($N_{HCO^+}$)}           & \textbf{Log($N_{CS}$)}             \\ \hline
40.0  & AGN    & $4.40_{0.83}^{0.57} $ & $240.19_{148.66}^{117.72} $ & $94.33_{31.79}^{36.44} $ & $16.84_{1.21}^{0.52} $ & $15.85_{1.06}^{0.41} $ & $15.74_{0.95}^{0.39} $ \\ \hline
40.0  & E Knot & $5.31_{0.31}^{0.25} $ & $125.52_{49.54}^{209.49} $  & $96.99_{32.42}^{33.98} $ & $16.96_{0.45}^{0.73} $ & $15.50_{0.25}^{0.23} $ & $15.73_{0.3}^{0.3} $   \\ \hline
40.0  & W Knot & $5.12_{0.23}^{0.35} $ & $172.63_{94.41}^{165.44} $  & $95.87_{31.85}^{34.69} $ & $16.88_{0.4}^{0.72} $  & $15.40_{0.27}^{0.22} $ & $15.59_{0.31}^{0.3} $  \\ \hline
40.0  & CND-N  & $4.74_{0.91}^{0.45} $ & $249.51_{148.66}^{103.06} $ & $95.77_{33.04}^{36.59} $ & $16.77_{0.66}^{0.76} $ & $15.43_{0.54}^{0.47} $ & $15.32_{0.52}^{0.47} $ \\ \hline
40.0  & CND-S  & $4.05_{0.5}^{0.46} $  & $234.44_{128.57}^{116.56} $ & $94.73_{31.75}^{36.32} $ & $17.15_{0.58}^{0.47} $ & $15.22_{0.51}^{0.49} $ & $15.49_{0.61}^{0.53} $ \\ \hline
40.0  & NSB-a  & $5.02_{0.17}^{0.23} $ & $260.80_{125.93}^{97.43} $  & $83.62_{24.1}^{38.03} $  & $16.23_{0.31}^{0.19} $ & $15.90_{0.29}^{0.22} $ & $16.13_{0.31}^{0.18} $ \\ \hline
40.0  & SSB-a  & $4.74_{0.31}^{0.27} $ & $235.52_{112.03}^{117.53} $ & $79.88_{21.31}^{40.57} $ & $16.48_{0.37}^{0.22} $ & $16.14_{0.36}^{0.22} $ & $16.08_{0.49}^{0.22} $ \\ \hline
40.0  & SSB-b  & $4.49_{0.45}^{0.46} $ & $264.01_{121.79}^{93.94} $  & $94.68_{32.05}^{32.82} $ & $16.40_{0.64}^{0.32} $ & $15.74_{0.59}^{0.33} $ & $15.98_{0.47}^{0.29} $ \\ \hline
40.0  & SSB-c  & $4.33_{0.59}^{0.46} $ & $259.91_{114.64}^{95.31} $  & $96.63_{32.73}^{34.81} $ & $16.16_{1.06}^{0.64} $ & $15.52_{1.01}^{0.66} $ & $15.43_{1.07}^{0.64} $ \\ \hline
100.0 & AGN    & $3.89_{0.6}^{0.8} $   & $241.12_{121.22}^{107.33} $ & $98.56_{33.84}^{34.92} $ & $16.11_{1.24}^{0.89} $ & $14.92_{1.1}^{0.8} $   & $14.91_{1.14}^{0.74} $ \\ \hline
100.0 & E Knot & $3.92_{0.62}^{0.77} $ & $234.73_{116.89}^{110.79} $ & $98.56_{33.1}^{34.22} $  & $16.36_{1.18}^{0.91} $ & $15.08_{1.0}^{0.79} $  & $14.55_{1.02}^{0.8} $  \\ \hline
100.0 & W Knot & $3.98_{0.63}^{0.83} $ & $259.74_{133.26}^{97.66} $  & $99.56_{33.29}^{33.05} $ & $16.44_{1.32}^{0.87} $ & $15.14_{0.93}^{0.81} $ & $14.85_{0.97}^{0.83} $ \\ \hline
100.0 & CND-N  & $3.96_{0.62}^{0.77} $ & $253.54_{126.31}^{98.41} $  & $98.15_{33.47}^{35.72} $ & $16.23_{1.29}^{0.87} $ & $15.06_{1.11}^{0.79} $ & $15.10_{1.14}^{0.79} $ \\ \hline
100.0 & CND-S  & $4.19_{0.7}^{0.6} $   & $251.54_{137.77}^{105.76} $ & $94.61_{31.48}^{36.2} $  & $16.70_{1.25}^{0.55} $ & $15.45_{0.96}^{0.48} $ & $15.65_{1.0}^{0.5} $   \\ \hline
100.0 & SSB-a  & $4.76_{1.11}^{1.33} $ & $227.38_{126.33}^{123.77} $ & $94.84_{30.33}^{37.0} $  & $15.60_{1.45}^{1.16} $ & $15.12_{1.16}^{0.99} $ & $14.32_{1.28}^{1.36} $ \\ \hline
\end{tabular}

\label{tab:Best_parameters}
\end{table*}

\subsection{Non-LTE analysis}
\label{subsec:Non-LTE}

In order to understand how the physical conditions of the CND regions and the SB regions relate to the possible tracers discussed in this paper, we compared the data with a grid of Large Velocity Gradient (LVG) models. We used {\tt RADEX} for this task via the {\tt SpectralRadex} python package\footnote{\url{https://spectralradex.readthedocs.io}}.

{\tt RADEX} is a non-LTE radiative transfer code developed by \cite{Van_der_Tak_2007}, which uses the escape probability method \citep{1960_Sobolev}. We assumed the sub-regions could be approximated as uniform spheres and computed the line intensities of each molecule for different physical parameters. To do this, we took collisional data between each molecule and H$_2$ from the LAMDA database \citep{Schoier_2005}.

We then ran a grid of physical parameters for each molecule. In this grid, we varied the  physical parameters \textbf{over the ranges summarised in Table \ref{tab:Grid_Range}}. 
 We based the gas density and kinetic temperature ranges on previous studies of NGC 1068 \citep{Viti2014,Izumi2016}, while the column density ranges were taken from the LTE analysis in Sect.~\ref{subsec:LTE Analysis}. In addition to these parameters, we tried three line widths that cover ranges of typical line widths observed in the SB regions (50 km s$^{-1}$) and the CND regions (150 km s$^{-1}$) \citep{Scourfield_2020}.



We used the resulting line intensities to compute the ratios that we have explored in this paper, namely: HCN(1-0)/HCO$^+$(1-0), HCN(4-3)/HCO$^+$(4-3), HCN(4-3)/CS(7-6), and HCN(4-3)/CS(2-1). We then fit these modelled ratios to their observed values by MCMC sampling the $\chi^2$ distribution of the parameters. For all parameter combinations, we computed the $\chi^2$ statistic:

\begin{equation}
    \chi^2=\sum_{i=1}^4(R_o-R_m)/\sigma^2,
\end{equation}
in order to compare goodness of fit. $R_o$ and $R_m$ are the observed and modelled ratios, respectively, and $\sigma$ is the uncertainty in the observed ratio propagated from the uncertainty in the individual intensities. Assuming a uniform prior, sampling this distribution gives the posterior probability distribution of the parameters.

Our MCMC sampler uses 12 walkers, each performing 100000 steps, to sample possible parameter combinations. We did this for all combinations of regions and resolutions with at least two of the investigated ratios. The joint posterior distributions of temperature and density are shown in Figure \ref{fig:grid_Chi} for each of the available regions at the highest resolution ($\sim$40 pc scale). Table \ref{tab:Best_parameters} summarises the parameters of the MCMC fitting of the \texttt{RADEX} models over the entire parameter space for nine of the ten regions studied in \cite{Scourfield_2020}, with NSB-c being excluded due to a lack of observed ratios, as stated in Section \ref{subsec:obscompare}. Neither the CND nor the SB ring regions can be reasonably constrained in the temperature or density planes, at least with regards to showing distinguishing qualities within these parameters. The only two regions in the CND where strong constraints on the gas density can be drawn are the east and west knot regions  ($5.31_{0.31}^{0.25}$ and $5.12_{0.23}^{0.35}$, respectively). Similarly the SB ring region, NSB-a, has a  relatively highly constrained density of $5.02_{0.23}^{0.17}$. The impact of resolution when it comes to the determination of temperature, density, and linewidth, with regards to each region, is negligible as a result of the fact that these parameters are not reasonably constrained at any resolution. However, it must be noted that the density becomes extensively less constrained for all regions as the resolution is diminished; this can be seen in  Fig. \ref{fig:100pc_Temp_Dens}. 
Despite the lack of strong constraints, it is clear from this that  the best-fit gas densities and temperatures are  generally quite similar between the CND and the SB regions. This strongly suggests that we cannot link the difference in ratios between SB and CND regions to either of these quantities. While the results for the temperature are relatively unconstrained, it is notable that the median values (at high resolution) for the SB region appear to be higher than those located in the CND, particularly those in the CND-N and CND-S. However, while typically the CND regions, by virtue of being located closer to the AGN, would be expected to have higher temperatures, it is possible to imagine a scenario of high SB activity  that could lead to average temperatures of the order of $\sim$300K (typical of galactic hot cores, e.g. \cite{Kaufman_1998}). On the other hand, we also note that there is significant uncertainty in these obtained temperature values.


Figures \ref{fig:Range_N(HCN)} to \ref{fig:Range_N(CS)} display the 1 sigma ranges of the posterior distributions for the column densities of HCN, HCO$^+$, and CS as a result of the $\sim$40 pc and $\sim$100 pc data. While there are some general trends towards  lower (for HCN) or higher (for HCO$^+$ and CS) column densities in the SB ring regions with respect to the CND regions, especially at the 40 pc scale, there is a significant overlap between the regions.



To account for the fact that the total H$_2$ column density varies across individual regions, and thus may be obscuring an underlying HCN enhancement, we computed the posterior probability distributions of the ratios of the column densities. We did this by taking the ratio of the chains of column density samples for each region and resolution. 
The resulting 16th and 84th percentile limits of the three ratio distributions are shown in Figure \ref{fig:grid_Abundance_ratios}. What we observe at our highest resolution is that the best fitting \texttt{RADEX} models typically have higher abundances of HCN relative to both HCO$^+$ and CS within the AGN-dominated CND regions when compared to the best fitting models of the SB-dominated SB ring regions.
This distinction is maintained somewhat in the lower-resolution ($\sim$100 pc scale) data for HCN relative to HCO$^+$, but not for HCN with respect to CS.
The abundances of CS and HCO$^+$ appear to cover relatively similar ranges, regardless of the region observed.

It is therefore the underlying ratio of the HCN/HCO$^+$ and the HCN/CS abundances \textbf{that distinguish the process characteristics of an AGN-dominated region or galaxy versus that of an SB-dominated region or galaxy}. Thus, line ratios are only useful insofar as they probe the column density differences, and therefore  multi-line multi-species observations coupled with chemical modelling will always be a more powerful, if not essential, method to discriminate and characterise different types of galaxies, while minimising degeneracy in radiative transfer modelling.

\section{Summary and conclusion}
\label{sec:Conclusion}

By taking NGC 1068 as a `laboratory',  we have investigated whether molecular line ratios are useful tracers of AGN-dominated gas. More importantly we determined  the origin of the differences in such ratios across different types of gas.   Our main conclusions are the following:

\begin{itemize}
    \item For the HCN(1-0)/HCO$^+$(1-0) and HCN(4-3)/CS(7-6) ratios, we observe an overlap in the observed ranges of these ratios between AGN-dominated and SB-dominated regions. For HCN(4-3)/CS(7-6) specifically, we notice that the range of values observed in each of the two types of regions overlapped significantly even at the highest spatial resolution.
    \item The HCN(1-0)/HCO$^+$(1-0) ratio first proposed in \cite{Kohno_2005} and then put into question by \cite{Privon_2020} does appear to show an enhancement in the CND with respect to the SB ring regions for the majority of sub-regions; however, we found an `outlier' in at least one SB region, (NSB-c).
    \item HCN(4-3)/HCO$^+$(4-3), on the other hand, 
    seems to be distinctively higher in all the regions closer to the AGN position, at least at higher spatial resolutions. This distinction is even maintained when the ratio is viewed at lower-resolution scales, although the separation does decrease. It is important to note, however, that for NSB-c the HCN(4-3) and HCO$^+$(4-3) lines are too weak, and thus we cannot confirm whether this ratio in this region conforms to the trend stated above.
    \item When observed at high spatial resolution, we propose a new ratio as a reliable tracer of AGN activity: the HCN(4-3)/CS(2-1).  
    
    \item We investigated the origin of the differences in ratios and found that differences in gas densities and temperatures are not the cause of the differentiation. Upon computing the relative fractional abundances of HCN, HCO$^+$, and CS,  we in fact determined that these differences are a consequence of differences in chemical abundances. Hence, it is essential to consider the chemistry of the species (i.e. which chemical processes lead to such abundances) when drawing conclusions from radiative transfer calculations \citep{Viti_2017}.
    \item As also noted in  \cite{2014_Garcia_Burillo} and \cite{Viti2014},  the elevated HCN(1-0)/HCO$^+$(1-0) and HCN(4-3)/HCO$^+$(4-3) ratios observed throughout the CND can be attributed to the molecular outflow observed in this region. As such, the contribution of mechanical heating and possible shock chemistry in the CND could be possible contributors to the observed HCN enhancement \citep{2017Kelly,2022Huang}.
\end{itemize}

\textbf{In conclusion, we have attempted to  evaluate the effectiveness of selected molecular line ratios as tracers of AGN versus SB activity in the galaxy NGC 1068. We have found that previously proposed ratios such as HCN(1-0)/HCO+(1-0) and HCN(4-3)/CS(7-6) do not appear to be sufficient  regardless of the spatial scale investigated, whereas HCN(4-3)/HCO+(4-3) HCN(4-3)/CS(2-1) may show promise at certain resolutions but follow up studies, especially of other nearby galaxies, is required in order to confirm our findings. Finally, we find that the observed variances in these ratios are a result of variances in the chemical abundances and hence chemical modelling is necessary in order to determine the origin of the differences in ratios.}
\section*{Acknowledgements}
This work is part of a project that has received funding from the European Research Council (ERC) under the European Union’s Horizon 2020 research and innovation programme MOPPEX 833460. SGB acknowledges support from the research project PID2019-106027GA-C44 of the Spanish Ministerio de Ciencia e Innovación. This paper makes use of the following ALMA data: (project-ID: 2013.1.00055.S; PI: S. Garc\'{ı}a-Burillo). The published data displayed in the following papers were also used in this study: \cite{Sanchez_Garcia_2022}, \cite{GarciaBurillo_S_2008}, \cite{2014_Garcia_Burillo}, \cite{Viti2014}, \cite{Perez_Beaupuits1009}, \cite{Tan2018}, \cite{Scourfield_2020}, \cite{Takano_2014}, \cite{Tacconi1997}, \cite{Nakajima2015}, \cite{Bayet_2009}. The authors would like to thank the anonymous referee for constructive comments that improved the original version of the paper.

%
%
\bibliographystyle{aa}
\bibliography{refs} 

\onecolumn
\clearpage
\appendix

\section{Additional figures}
\begin{figure}[!h]

\centering
\includegraphics[width=0.75\textwidth]{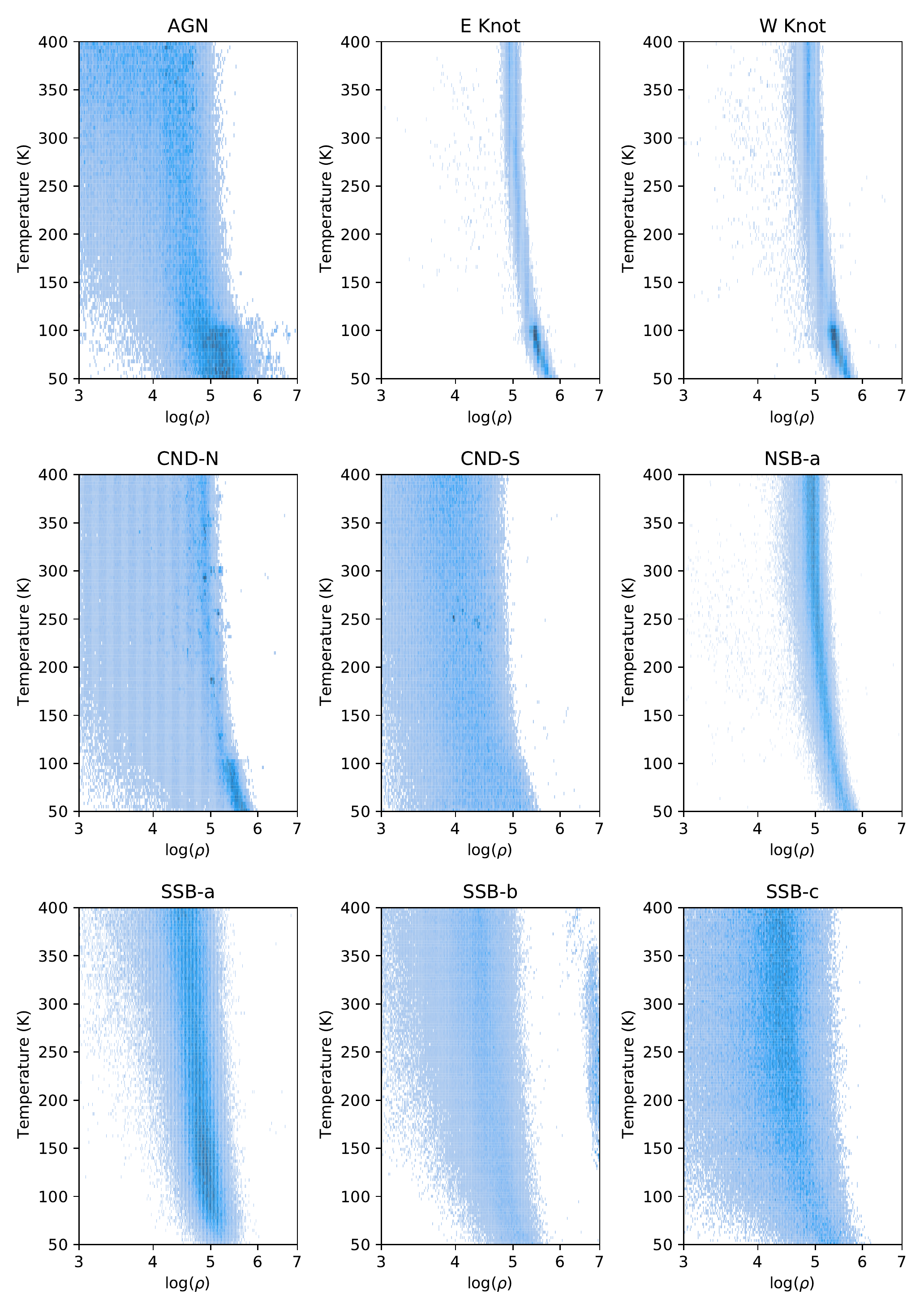}

\caption{Resulting joint posterior distributions for temperature and density as produced for the high resolution ($\sim40$ pc scale data) for the nine regions that have enough ratios to fit. The temperature and density values shown in Table \ref{tab:Best_parameters} are from distributions visualised in this plot. It must be noted that the median values of these distributions, as shown in Table \ref{tab:Best_parameters}, do not always correspond to the most densely populated areas of the corresponding plot. This is due to the extent of the posterior distribution in the temperature plane.}
\label{fig:grid_Chi}
\end{figure}

\begin{figure*}

\centering
\includegraphics[width=0.8\textwidth]{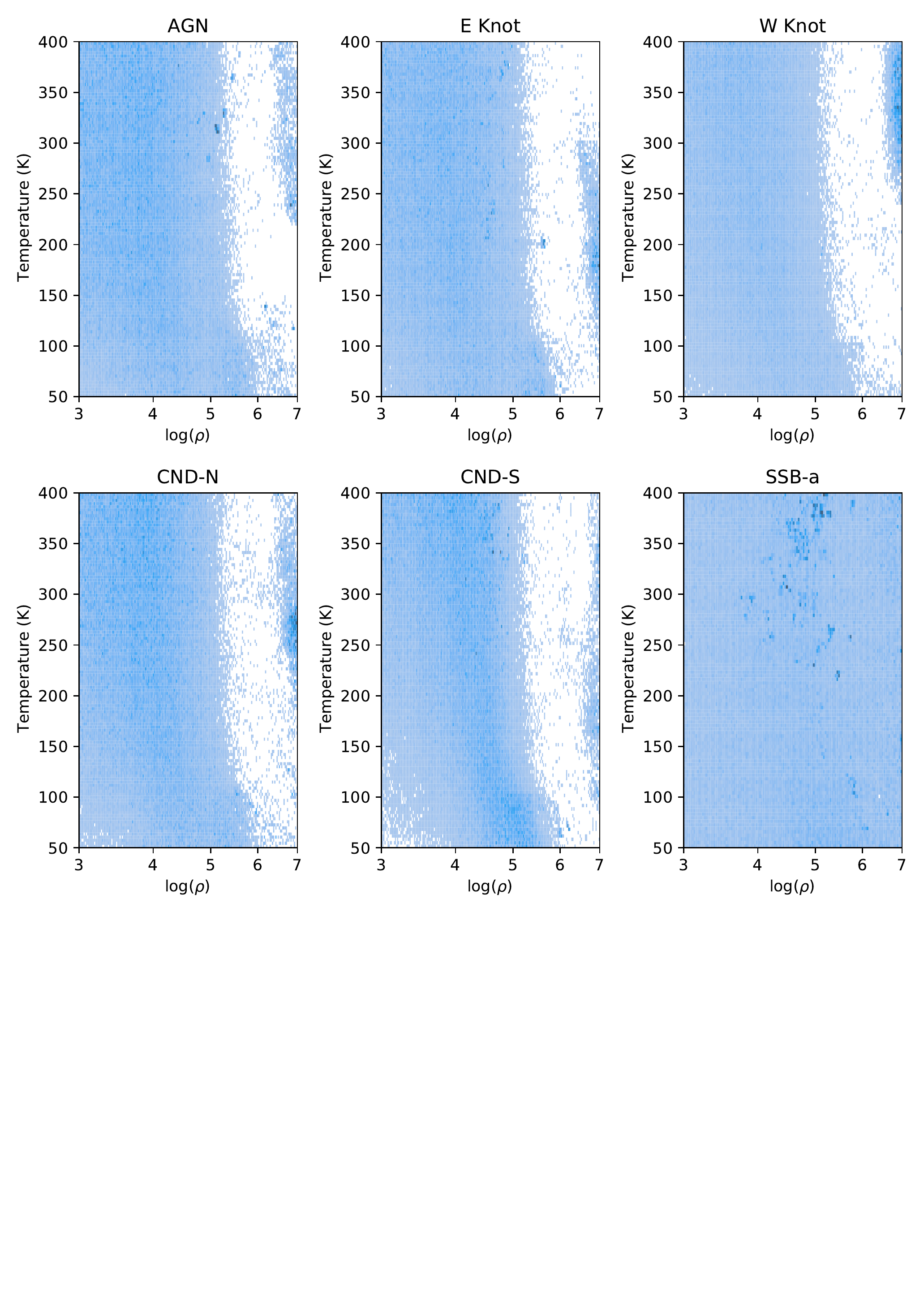}

\caption{Resulting joint posterior distributions for temperature and density as produced for the lower resolution ($\sim100$ pc scale data) for the six regions that have enough ratios to fit, at this lower resolution.}
\label{fig:100pc_Temp_Dens}
\end{figure*}
\twocolumn
\begin{figure}

\centering
\includegraphics[width=0.4\textwidth]{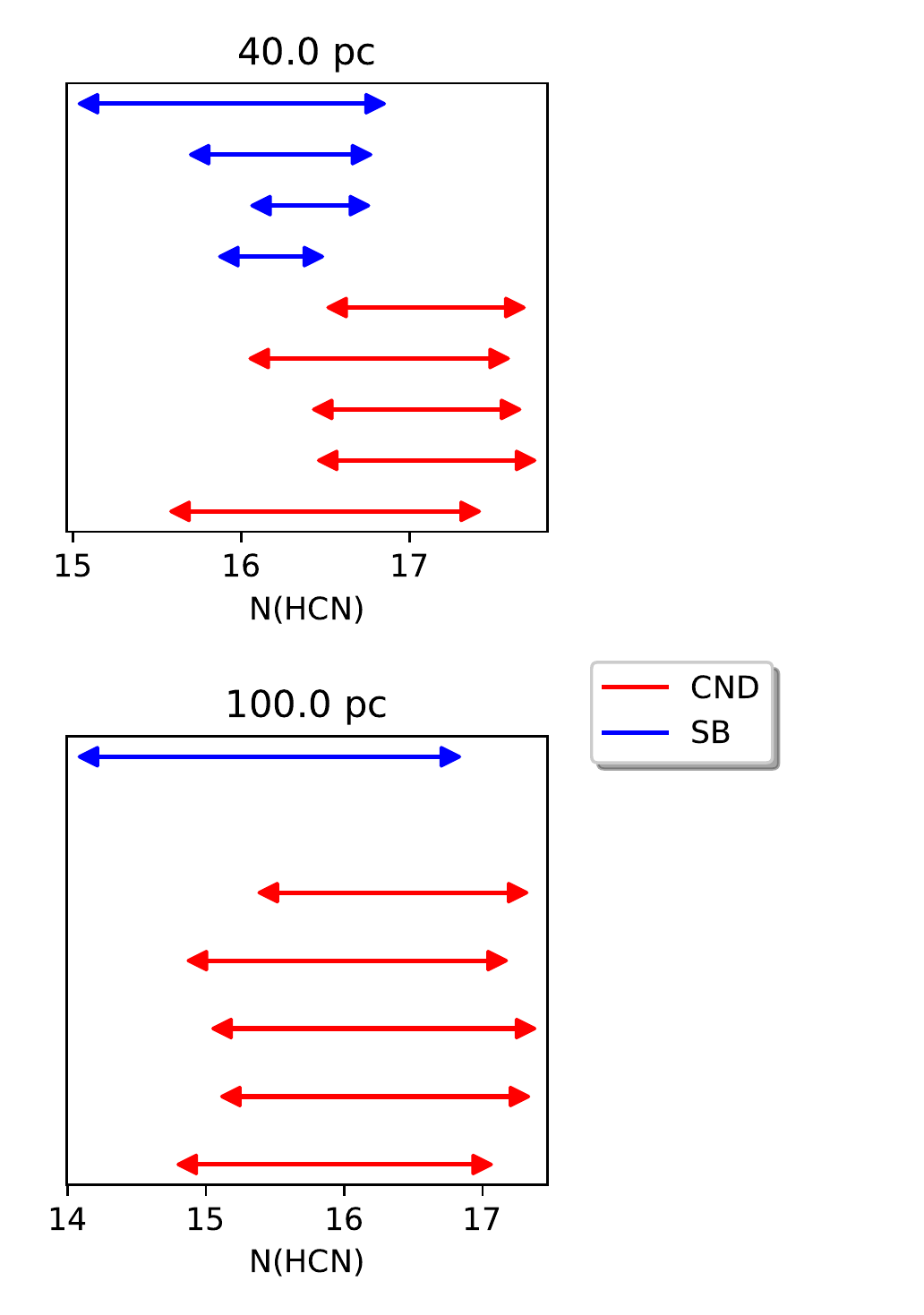}

\caption{16th to 84th percentile range of the MCMC for the column density of HCN, as found from fitting the observed ratios to those produced by \texttt{RADEX} models.}
\label{fig:Range_N(HCN)}
\end{figure}

\begin{figure}

\centering
\includegraphics[width=0.4\textwidth]{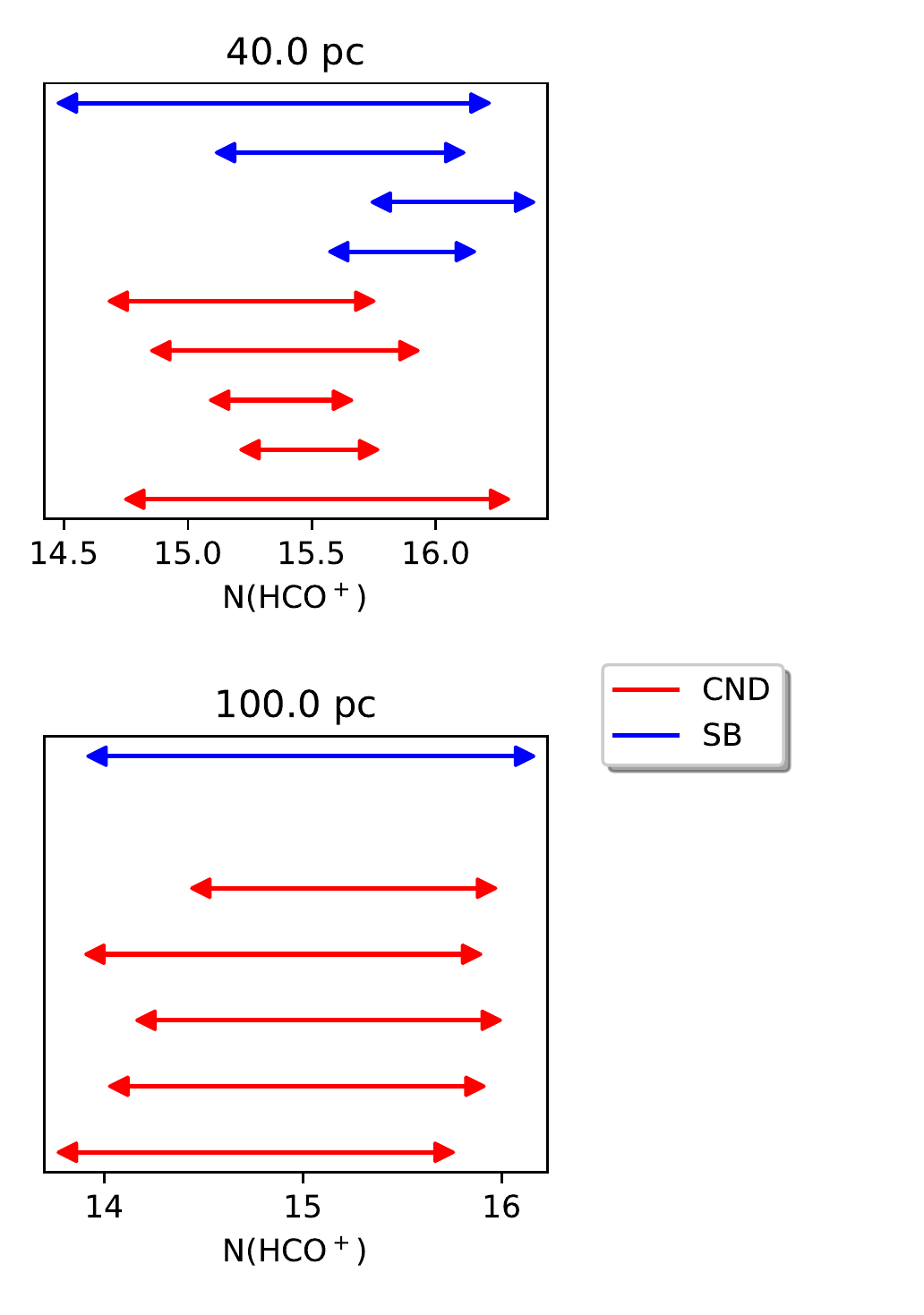}

\caption{16th to 84th percentile range of the MCMC for the column density of HCO$^+$, as found from fitting the observed ratios to those produced by \texttt{RADEX} models.}
\label{fig:Range_N(HCOP)}
\end{figure}

\begin{figure}

\centering
\includegraphics[width=0.4\textwidth]{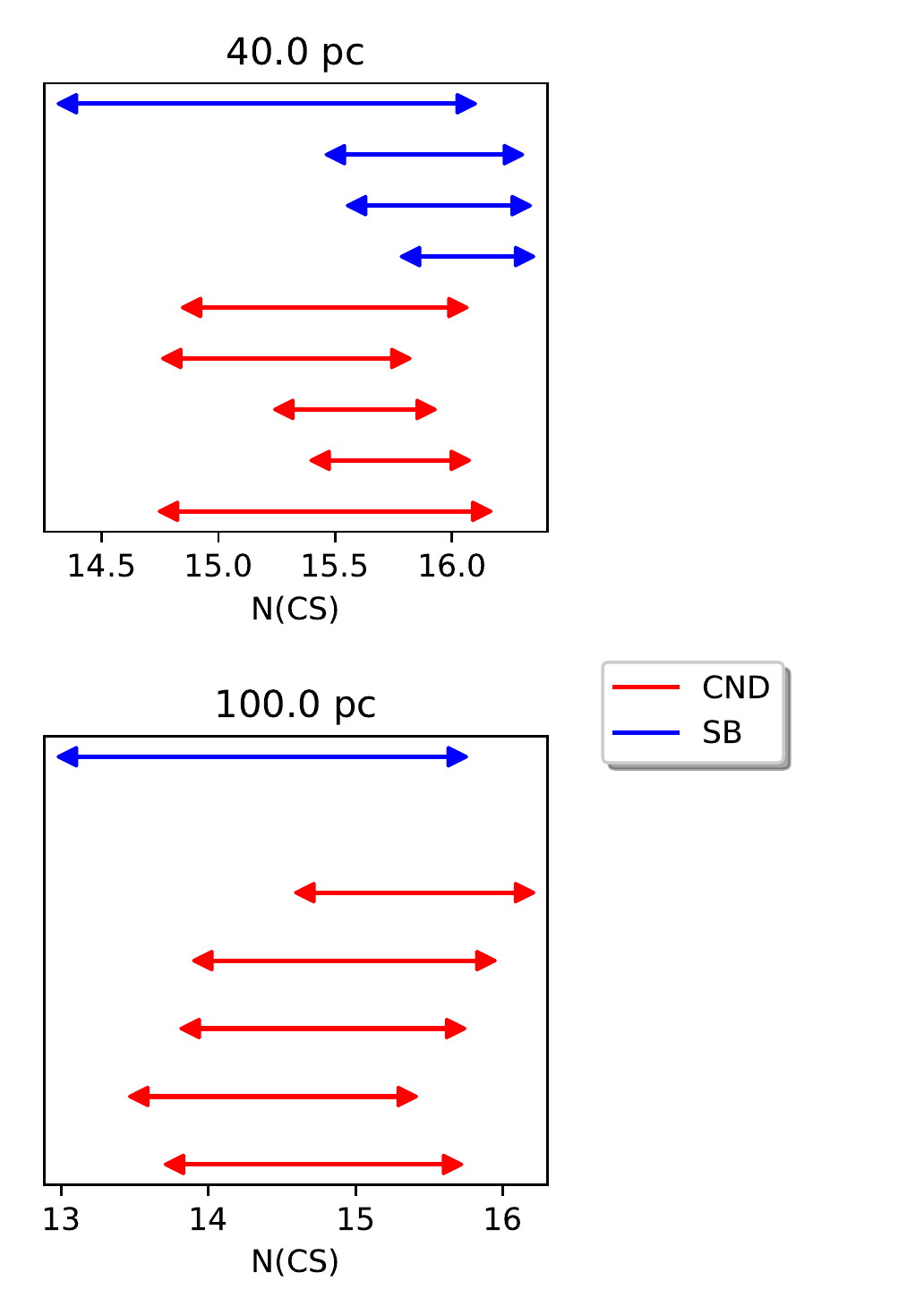}

\caption{16th to 84th percentile range of the MCMC for the column density of CS, as found from fitting the observed ratios to those produced by \texttt{RADEX} models.}
\label{fig:Range_N(CS)}
\end{figure}

\begin{figure*}

\centering
\includegraphics[width=0.8\textwidth]{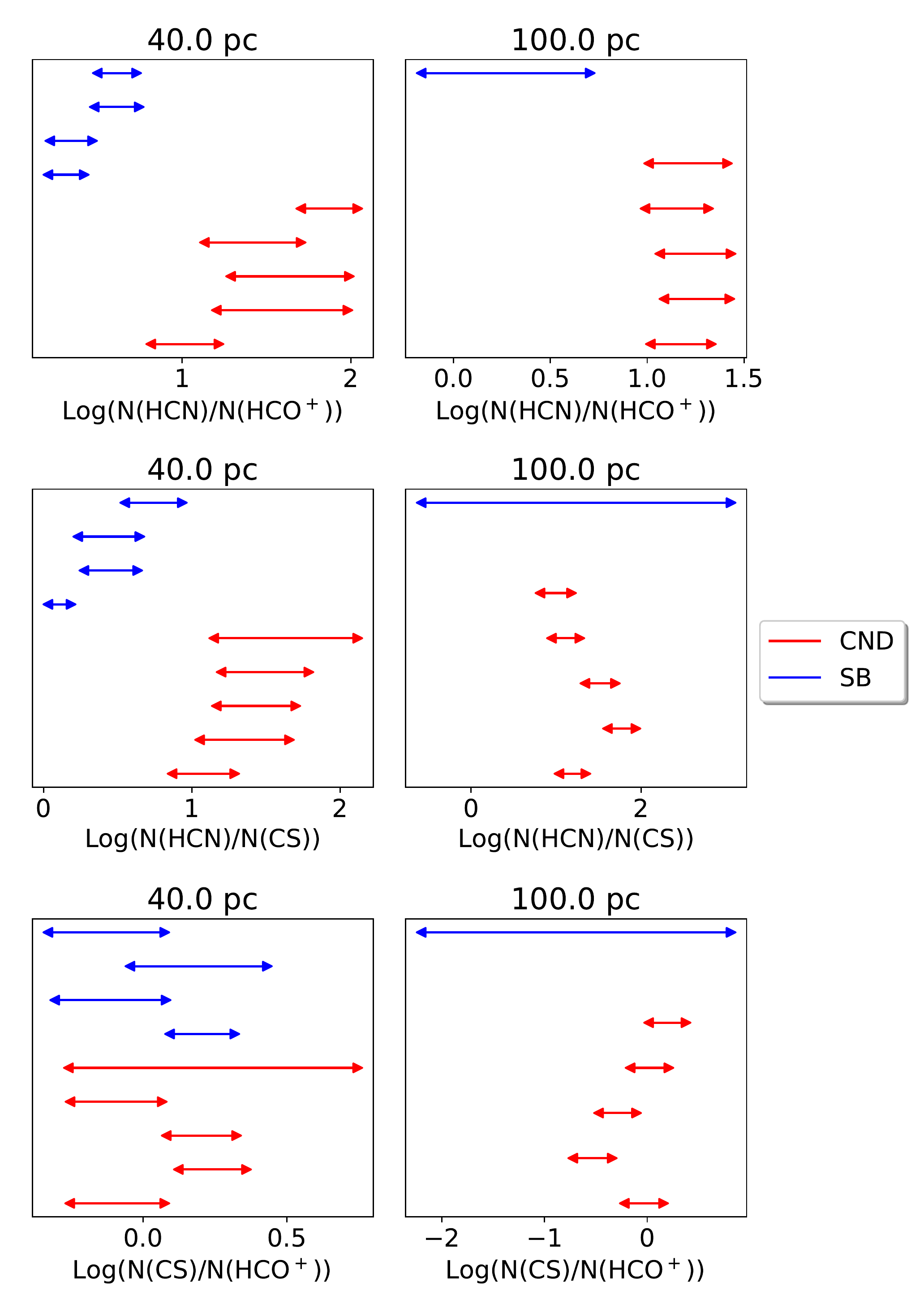}

\caption{Produced relative fractional abundances of HCN, HCO$^+$, and CS across the various regions in NGC 1068 at both the 40 pc and 100 pc scale.}
\label{fig:grid_Abundance_ratios}
\end{figure*}

\FloatBarrier
\onecolumn
\section{Additional moment 0 maps}
\subsection{HCN(4-3)}
\label{appsubsec:HCN_43}
\begin{figure*}[h!]
\centering
\includegraphics[width = 0.9\textwidth]{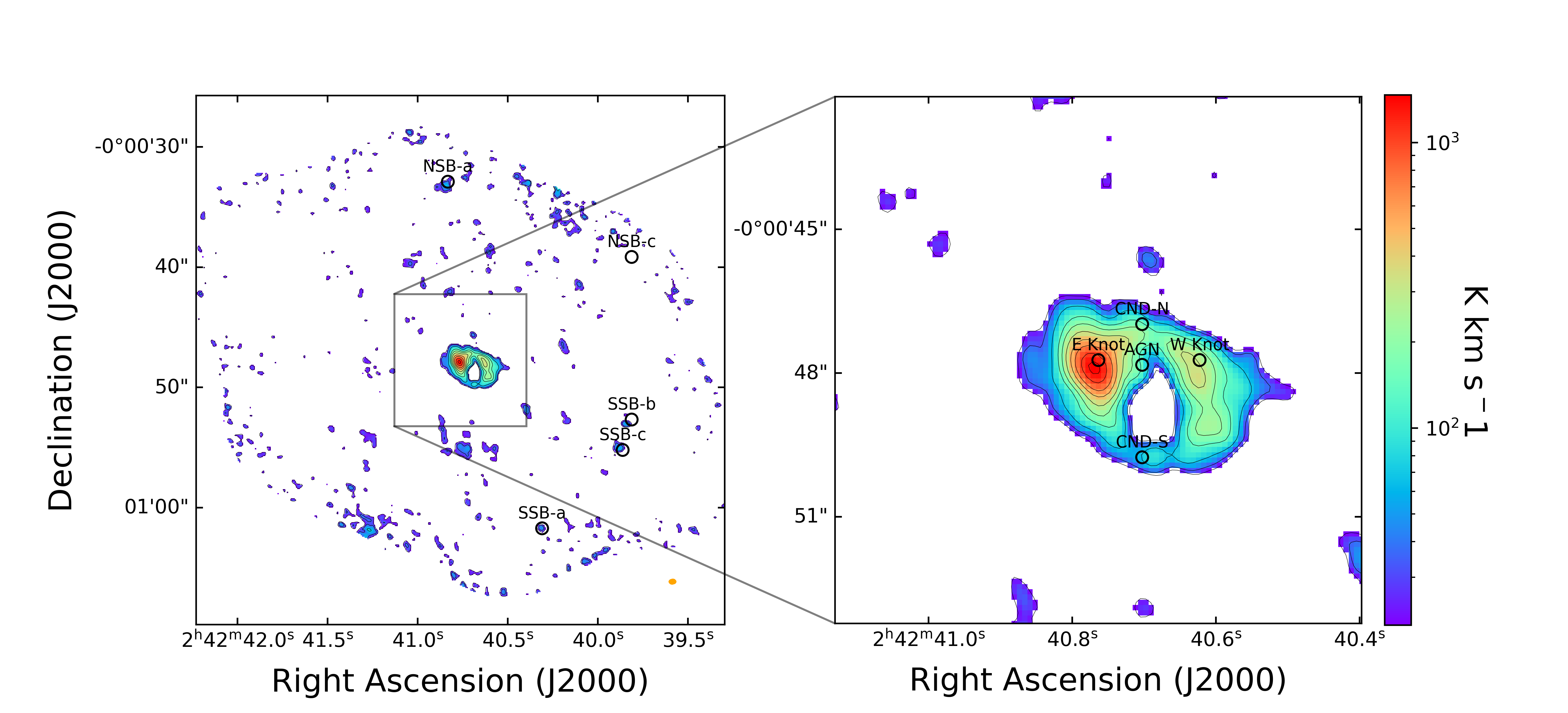}
\caption{HCN J(4-3) velocity-integrated moment map. The beam size is shown by the orange ellipse. The lowest contour displayed is 3$\sigma$, where $\sigma=$6.8 K km s$-1$, with the following contours corresponding to 5$\sigma$, 10$\sigma$, 20$\sigma$, 30$\sigma$, 45$\sigma$, 70$\sigma$, 100$\sigma$, and 150$\sigma$. }
\label{fig:HCN_43}
\end{figure*}

\subsection{HCO$^{+}$(4-3)}
\label{appsubsec:HCOP_43}
\begin{figure*}[!htb]
\centering
\includegraphics[width = 0.9\textwidth]{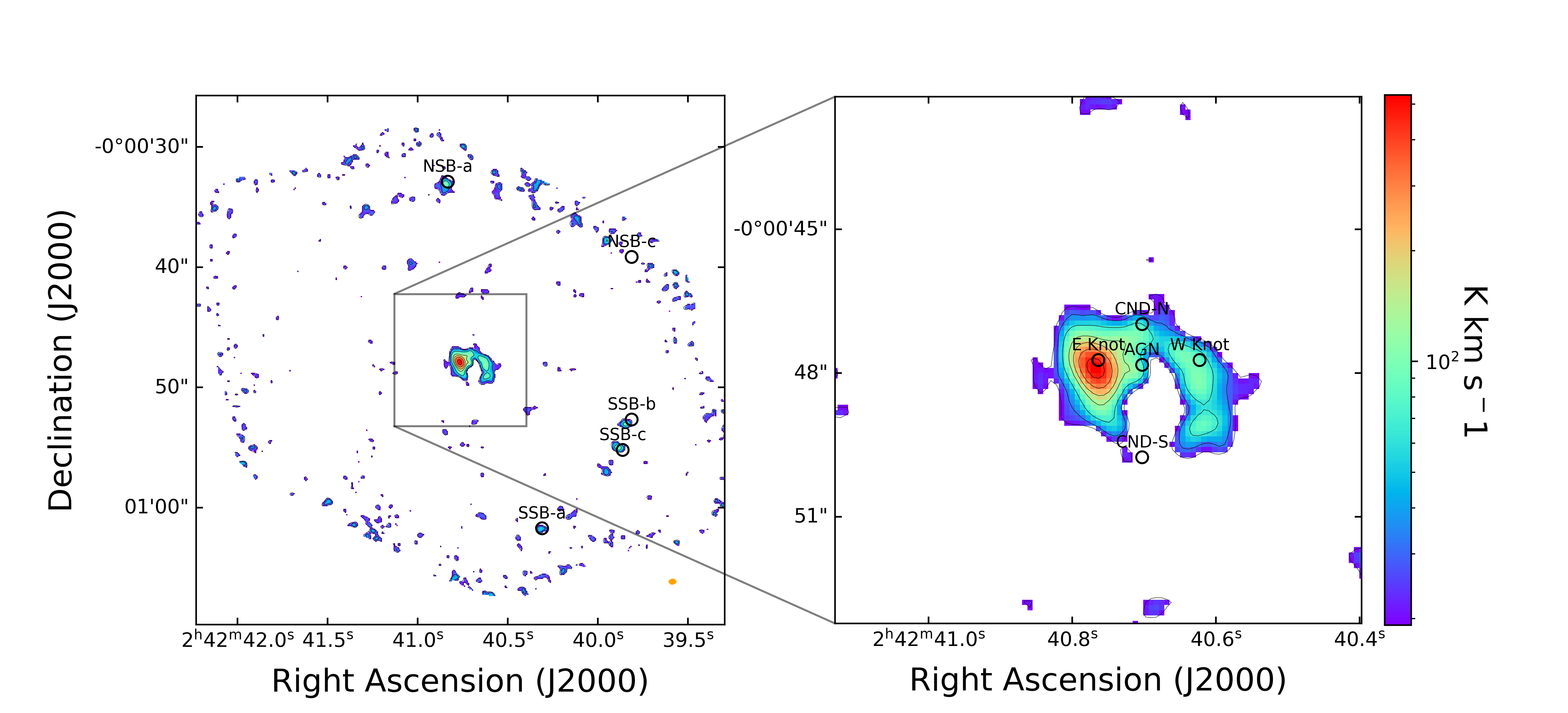}
\caption{HCO$^{+}$ J(4-3) velocity-integrated moment map. The beam size is shown by the orange ellipse. The lowest contour displayed is 3$\sigma$, where $\sigma=$6.4 K km s$-1$, with the following contours corresponding to 5$\sigma$, 10$\sigma$, 20$\sigma$, 30$\sigma$, 45$\sigma$, 70$\sigma$, 100$\sigma$, and 150$\sigma$. }
\label{fig:HCOP_43}
\end{figure*} 
\vfill
\clearpage
\end{document}